# Optimal distance- and time-dependent area-based pricing with the Network Fundamental Diagram

Ziyuan Gu[a], Sajjad Shafiei[b], Zhiyuan Liu[c], Meead Saberi[a,*]

[a] *School of Civil and Environmental Engineering, University of New South Wales, NSW 2052, Australia*
[b] *Institute of Transport Studies, Department of Civil Engineering, Monash University, VIC 3800, Australia*
[a] *Jiangsu Key Laboratory of Urban ITS, Jiangsu Province Collaborative Innovation Center of Modern Urban Traffic Technologies, School of Transportation, Southeast University, Nanjing 210096, China*

*\* Corresponding author: meead.saberi@unsw.edu.au*

**Abstract**

Given the efficiency and equity concerns of a cordon toll, this paper proposes a few alternative distance-dependent area-based pricing models for a large-scale dynamic traffic network. We use the Network Fundamental Diagram (NFD) to monitor the network traffic state over time and consider different trip lengths in the toll calculation. The first model is a distance toll that is linearly related to the distance traveled within the cordon. The second model is an improved joint distance and time toll (JDTT) whereby users are charged jointly in proportion to the distance traveled and time spent within the cordon. The third model is a further improved joint distance and delay toll (JDDT) which replaces the time toll in the JDTT with a delay toll component. To solve the optimal toll level problem, we develop a simulation-based optimization (SBO) framework. Specifically, we propose a simultaneous approach and a sequential approach, respectively, based on the proportional-integral (PI) feedback controller to iteratively adjust the JDTT and JDDT, and use a calibrated large-scale simulation-based dynamic traffic assignment (DTA) model of Melbourne, Australia to evaluate the network performance under different pricing scenarios. While the framework is developed for static pricing, we show that it can be easily extended to solve time-dependent pricing by using multiple PI controllers. Results show that although the distance toll keeps the network from entering the congested regime of the NFD, it naturally drives users into the shortest paths within the cordon resulting in an uneven distribution of congestion. This is reflected by a large clockwise hysteresis loop in the NFD. In contrast, both the JDTT and JDDT reduce the size of the hysteresis loop while achieving the same control objective. We further conduct multiple simulation runs with different random seed numbers to demonstrate the effectiveness of different pricing models against simulation stochasticity. However, we postulate that the feedback control is not applicable with guaranteed convergence if the periphery of the cordon area becomes highly congested or gridlocked.

## 1. Introduction

Road user pricing has been studied extensively both in practice and in theory as an effective means of mitigating urban traffic congestion. Back in 1975, Singapore successfully launched the world's first pricing scheme called the area licensing scheme. Following Singapore's success, several other pricing schemes have been implemented worldwide, such as in London, Stockholm, and Milan. See Gu et al. (2018) for a comprehensive overview. More recently, instead of looking into cordon and zonal schemes, a number of advanced pricing schemes have been proposed and implemented thanks to the development of various pricing technologies. For example, (i) the Move NY Plan in New York City which aims to charge taxis based on both distance and time[1]; (ii) Singapore's ERP 2.0 (from 2020) which uses satellites to charge vehicles based on distance[2] traveled; (iii) the opt-in user-pays system OReGo in Oregon, USA which is a state-wide distance-based charge[3]; (iv) a per-km charging system considered by the Metro Vancouver Independent Mobility Pricing Commission (Lee, 2018); and (v) a joint distance- and cordon-based scheme trialed in Melbourne (Transurban, 2016). Interested readers may refer to De Palma and Lindsey (2011) for an overview of different technologies applied to pricing.

In theory, since the seminal studies by Pigou (1920) and Knight (1924), a variety of models have been proposed for designing an optimal pricing system. These models, however, cannot be easily applied to a large-scale dynamic traffic network because of their demanding requirement of highly detailed information of each origin-destination (OD) pair and each individual link in the network. See Yang and Huang (2005) for a comprehensive overview. Recent advances in network traffic flow theory through the Network Fundamental Diagram (NFD) or Macroscopic Fundamental Diagram (MFD) (Geroliminis and Daganzo, 2008; Mahmassani et al., 2013) have established a new branch of pricing theory that largely facilitates the study, design, and implementation of large-scale pricing models. Nevertheless, studies so far are mainly limited to the cordon-based regime which, although being simple, suffers from inefficiency and inequity. Therefore, this paper aims to propose a few alternative distance-dependent area-based pricing models for large-scale dynamic traffic networks. To find the optimal toll levels that keep the network from entering the congested regime of the NFD, we develop a simulation-based optimization (SBO) framework combining the NFD and a simulation-based dynamic traffic assignment (DTA) model.

*1.1. Literature review*

The first pricing model is known as first-best pricing or marginal-cost pricing (MCP), which was first applied to a single link (Li, 2002) and then to a general traffic network (Yang et al., 2004). Despite perfect theoretical basis, its practical applications are rather limited for two main reasons (Simoni et al., 2015; Yang and Huang, 2005): (i) charging all links in a network is infeasible and inefficient given the high operating cost and the poor public acceptance; and (ii) assuming an abstract and constant demand-supply relationship is invalid. As a result, a variety of second-best pricing regimes have been proposed in which only a subset of links is tolled. A widely practiced way to solve the second-best pricing problem is to formulate it as a bi-level optimization problem or, equivalently, a mathematical program with equilibrium constraints (MPEC) (Liu and McDonald, 1999; Liu et al., 2013; Liu et al., 2014; Meng et al., 2012; Verhoef, 2002; Yang and Zhang, 2003; Zhang and Yang, 2004). Solving the MPEC particularly for a large-scale dynamic traffic network remains challenging because of the required computational complexity. Time-dependent information of each OD pair and each individual link in the network is needed for formulating the optimization model. Often the resulting objective function is of an expensive-to-evaluate type that cannot be solved directly through any exact solution algorithm (Chen et al., 2016; Chen et al., 2014; Ekström et al., 2016; He et al., 2017).

---

[1] See https://nyc.streetsblog.org/2015/02/17/the-complete-guide-to-the-final-move-ny-plan/.
[2] See https://www.lta.gov.sg/apps/news/default.aspx?scr=yes&keyword=ERP2.
[3] See http://www.myorego.org/.

To develop an effective and efficient pricing model for large-scale dynamic traffic networks, an understanding of traffic dynamics at the network level is critical (Zheng et al., 2016). Macroscopic traffic flow relations for an urban traffic network were initially proposed by Godfrey (1969) followed by Daganzo (2007); Mahmassani et al. (1987); Olszewski et al. (1995). The formal demonstration of the existence of the NFD was shown only recently using field data from Yokohama, Japan (Geroliminis and Daganzo, 2008). Since then, the NFD has been widely studied for network-wide control and management (Geroliminis et al., 2013; Keyvan-Ekbatani et al., 2012; Ramezani et al., 2015) due to its macroscopic nature and favorable properties (see Geroliminis and Daganzo (2008) for further description). In particular, an innovative NFD-based approach to modeling congestion pricing has emerged.

Since macroscopic modeling does not involve detailed route choice, an urban traffic network governed by an NFD can be modeled as a simple queuing system (e.g. a bottleneck) represented by the cumulative arrival and exit curves. Congestion is interpreted as vehicles queuing behind the bottleneck while demand results from their endogenous scheduling preferences considering the discrepancy between the desired and actual arrival times (Arnott et al., 1990; Arnott et al., 1993; Vickrey, 1963). Geroliminis and Levinson (2009) first integrated the bottleneck model with the NFD to consider the capacity drop. Daganzo and Lehe (2015) further introduced trip length heterogeneity and proposed an optimal usage-based toll. A fundamental study on NFD-based pricing in large-scale real-world networks was only recently conducted where the NFD was used to describe the level of congestion at the network level (Zheng et al., 2012). The authors applied an integral (I-type) controller for toll adjustment within an agent-based simulation environment. The I-type controller was later extended to a proportional-integral (PI) controller whereby users' adaptation to pricing was considered (Zheng et al., 2016). A similar simulation framework based on the NFD was also proposed by Simoni et al. (2015) but without using feedback control theory. The authors developed the toll adjustment rule by integrating the NFD (and also the 3D-NFD) with the MCP.

*1.2. Objectives and contributions*

NFD-based pricing in large-scale networks has so far mainly focused on the cordon-based regime. Given that a cordon toll undercharges long journeys and overcharges short trips, a distance toll was considered as an alternative (May and Milne, 2000; Meng et al., 2012). Although theory on distance-based pricing is well established, it is largely constrained within the traditional second-best pricing framework. Investigating distance-dependent pricing using the NFD therefore becomes theoretically interesting and also practically promising.

The overall toll design problem consists of a toll level problem and a toll area problem (Ekström et al., 2012). In this paper, we assume a predefined toll area and only aim to solve the optimal toll level problem. Since we are solving an area-based pricing problem, the NFD becomes a natural entry point to the solution due to its macroscopic nature. Similar to the existing studies on NFD-based network-wide control and management (Aboudolas and Geroliminis, 2013; Keyvan-Ekbatani et al., 2012; Simoni et al., 2015; Zheng et al., 2016; Zheng et al., 2012), the objective is to price and keep the network from entering the congested regime of the NFD. This is because when the network becomes congested or gridlocked (i.e. the network density increases beyond the critical threshold), the outflow/trip completion rate significantly drops causing network unproductivity. We assume that the NFD does not change significantly when we implement different pricing strategies on the network.

As a further extension to Zheng et al. (2012) and Zheng et al. (2016), we aim to propose a few alternative distance-dependent area-based pricing models for large-scale dynamic traffic networks. While using the NFD to describe the level of congestion at the network level, we consider different trip lengths in the toll calculation. The first model is a distance toll that is linearly related to the distance traveled within the cordon. A limitation, however, is that it naturally drives users into the shortest paths within the cordon resulting in an uneven distribution of congestion. The increased spatial traffic heterogeneity results in clusters of congested links within the cordon (i.e. pockets of congestion) and therefore (i) affects the shape of the NFD and deteriorates the network performance (Mazloumian et al., 2010; Saberi and Mahmassani, 2012, 2013); and (ii) compromises the toll optimality (Simoni et al., 2015). To address the limitation of the distance toll, we propose a joint distance and time toll (JDTT) as the second model based on Liu et al. (2014). The JDTT is linearly related to both the distance traveled and time spent within the cordon. Since the time toll component tends to overcharge users (because a longer link typically requires more travel time despite being uncongested), we further propose a joint distance and delay toll (JDDT) as the third model whereby users pay partially in proportion to their travel delays. To solve the optimal toll level problem, we develop an SBO framework which resembles, to some extent, the trial-and-error method in Yang et al. (2004). Specifically, we apply a PI feedback controller for iterative toll adjustment and use a calibrated large-scale simulation-based DTA model of Melbourne, Australia to evaluate different pricing scenarios. Practical applications of this type of iterative pricing strategy can be found in Singapore (Liu et al., 2013) and San Francisco (Pierce and Shoup, 2013). While the framework is developed for static pricing, we show that it can be easily extended to solve time-dependent pricing by using multiple PI controllers. That is, we implement a PI controller in each tolling interval instead of using only one for the entire tolling period.

This paper aims to show that the distance toll generates an uneven distribution of congestion resulting in a large clockwise hysteresis loop in the NFD, and that both the JDTT and JDDT reduce, to some extent, this negative effect. Although imposing an additional time-/delay-based charge on users, the JDTT/JDDT does not overcharge but is a trade-off between the distance and time/delay toll components. In summary, this paper offers the following contributions:

i. We show that the distance toll naturally drives users into the shortest paths within the cordon resulting in an uneven distribution of congestion and hence creating a large clockwise hysteresis loop in the NFD.
ii. To reduce the increased spatial traffic heterogeneity resulting from the distance toll, we integrate a time toll component and a delay toll component, respectively, in the toll calculation resulting in two improved area-based pricing models: JDTT and JDDT.
iii. To solve the optimal toll level problem, we develop an SBO framework and propose a simultaneous approach and a sequential approach, respectively, based on the PI controller for iteratively adjusting the JDTT and JDDT.
iv. We postulate a prerequisite for applying the PI controller: the periphery of the cordon area should have sufficient capacity to accommodate the re-routed traffic. If the prerequisite is not satisfied, the pricing control may fail resulting in a paradox where the toll levels keep rising but users still enter the cordon.

The remainder of the paper is organized as follows. Section 2 proposes the methodological framework for solving the optimal toll level problem. Section 3 presents a case study where the network performance under different pricing scenarios is discussed and compared. Section 4 postulates a prerequisite for applying the feedback control. Section 5 concludes the paper and makes recommendations for future research.

## 2. Methodology

Under the assumption that the NFD does not change significantly due to the imposition of different tolls, we propose our methodological framework consisting of the following four parts. Section 2.1 describes the construction of multiple macroscopic traffic flow variables using simulation outputs. Section 2.2 introduces the route choice model that is used in simulation considering different tolls. Section 2.3 proposes three feedback control strategies based on the PI controller for iteratively adjusting the toll rates. Section 2.4 formulates the complete toll level problem and develops an SBO framework as the solution algorithm.

While the framework developed in Section 2.4 is for static pricing (i.e. using only one PI controller for the entire tolling period), it can be easily extended to solve time-dependent pricing by implementing a PI controller in each tolling interval. Therefore, the proposed pricing strategies can be applied under both static and time-dependent scenarios.

### 2.1. Macroscopic traffic flow relations

The NFD can be estimated using different methods (Leclercq et al., 2014). In this paper, we use the outputs of a simulation-based DTA model to measure the NFD. The average network density $K$ and the average network flow $Q$ are determined, respectively, using Eqs. (1) and (2) over 5-minute intervals throughout the simulation time window, which are essentially spatially weighted averages of all the individual links (Saberi et al., 2014a; Zockaie et al., 2014):

$$K = \frac{\sum_{i=1}^{n} k_i \cdot l_i \cdot n_i}{\sum_{i=1}^{n} l_i \cdot n_i} \tag{1}$$

$$Q = \frac{\sum_{i=1}^{n} q_i \cdot l_i \cdot n_i}{\sum_{i=1}^{n} l_i \cdot n_i} \tag{2}$$

where $k_i$ and $q_i$ are the average density and flow of link $i$ over the observation period, respectively; $l_i$ and $n_i$ are the length and number of lanes of link $i$, respectively; $n$ is the total number of links in the network.

Since the spatial traffic heterogeneity is a key determinant of the shape and scatter of the NFD (Buisson and Ladier, 2009; Mazloumian et al., 2010), we further introduce the spatial spread of density $\gamma$ representing how congestion is distributed within the cordon. By definition (Knoop and Hoogendoorn, 2013), it is estimated using Eq. (3) as the square root of the weighted variance of all link densities:

$$\gamma = \sqrt{\frac{\sum_{i=1}^{n} l_i \cdot n_i \cdot (k_i - K)^2}{\sum_{i=1}^{n} l_i \cdot n_i}} \tag{3}$$

The spatial spread of density naturally increases with a growing accumulation. That is, an increase in vehicles entering the cordon inevitably generates a higher spatial spread of density later in time as these vehicles continue their trips within the cordon. Given the correlation between the spatial spread of density and accumulation, the uneven distribution of congestion is better interpreted as positive deviations from the natural increment represented by the lower envelope in the spread-accumulation relationship (Simoni et al., 2015). By fitting a polynomial function $\gamma(k)$ to the lower envelope, an indicator $\sigma$ is obtained and termed the deviation from spread:

$$\sigma = \gamma - \gamma(k) \tag{4}$$

## 2.2. Route choice model with tolls

Similar to second-best pricing, NFD-based pricing can be formulated as a bi-level optimization problem. The upper level forms the objective function which aims to keep the maximum network density within the tolling period around the critical threshold identified from the NFD. The lower level is a dynamic traffic assignment model representing how users respond to pricing by adjusting their route choice.

Consider a network denoted by $G = (N, A)$ where $N$ is the set of nodes and $A$ is the set of directed links. Given a predefined toll area, network $G$ is partitioned into a pricing sub-network $\bar{G} = (\bar{N}, \bar{A})$ and an external sub-network $\tilde{G} = (\tilde{N}, \tilde{A})$ satisfying $\bar{G} + \tilde{G} = G$. For the ease of presentation, notations are summarized in Table 1.

**Table 1** A summary of notations.

| Notation | Explanation |
|---|---|
| $W$ | The set of OD pairs, i.e. $W = \{(o,d) | o \in O, d \in D\}$ where $O \subset N$ is the set of origins and $D \subset N$ is the set of destinations |
| $R^{od}$ | The set of paths between an OD pair $(o,d) \in W$ |
| $q^{od}(h)$ | The travel demand between an OD pair $(o,d) \in W$ during interval $h$ |
| $f_r^{od}(h)$ | The traffic flow assigned to path $r \in R^{od}$ between an OD pair $(o,d) \in W$ during interval $h$ |
| $t_a(h)$ | The average travel time on link $a \in A$ during interval $h$ |
| $l_a$ | The length of link $a$ |
| $\delta_{a,r}^{od}$ | $\delta_{a,r}^{od} = 1$ if path $r \in R^{od}$ contains link $a$, otherwise $\delta_{a,r}^{od} = 0$ |
| $N_T$ | The total number of intervals throughout the simulation time window |

We assume that the distance, time, and delay toll functions $\varphi(d)$, $\phi(t)$, and $\psi(c)$ are linearly related to the distance traveled, time spent, and delay experienced within sub-network $\bar{G}$, respectively:

$$\varphi(d) = \alpha \cdot d \tag{5}$$

$$\phi(t) = \beta_1 \cdot t \tag{6}$$

$$\psi(c) = \beta_2 \cdot c \tag{7}$$

where $\alpha$, $\beta_1$, and $\beta_2$ are positive distance, time, and delay toll rates to be determined, respectively, and $c$ is the difference between the actual and free-flow travel times. Let $l_r^{od}(h)$ and $t_r^{od}(h)$ denote the total distance traveled and total time spent within sub-network $\bar{G}$ for path $r \in R^{od}$ during interval $h$, respectively:

$$l_r^{od}(h) = l_r^{od} = \sum_{a \in \bar{A}} l_a \cdot \delta_{a,r}^{od}, \quad r \in R^{od}, (o,d) \in W, h \in (1, \dots, N_T) \tag{8}$$

$$t_r^{od}(h) = \sum_{a \in \bar{A}} t_a(h) \cdot \delta_{a,r}^{od}, \quad r \in R^{od}, (o,d) \in W, h \in (1, \dots, N_T) \tag{9}$$

By integrating Eqs. (5-9), we obtain the distance, time, and delay toll components for path $r \in R^{od}$ during interval $h$, respectively:

$$\bar{\tau}_r^{od}(h) = \varphi\left(l_r^{od}(h)\right) = \alpha \cdot \sum_{a \in \bar{A}} l_a \cdot \delta_{a,r}^{od} \tag{10}$$

$$\tilde{\tau}_r^{od}(h) = \phi\left(t_r^{od}(h)\right) = \beta_1 \cdot \sum_{a \in A} t_a(h) \cdot \delta_{a,r}^{od} \tag{11}$$

$$\hat{\tau}_r^{od}(h) = \phi(t_r^{od}(h) - \bar{t}_r^{od}) = \beta_2 \cdot \sum_{a \in A} (t_a(h) - \bar{t}_a) \cdot \delta_{a,r}^{od} \tag{12}$$

where $\bar{t}_r^{od}$ and $\bar{t}_a$ are the free-flow travel times for path $r \in R^{od}$ and link $a \in \bar{A}$, respectively. Under the linearity assumption, we have the link-additive property of the distance, time, and delay toll components. With the distance toll, the generalized travel cost function $\bar{V}_r^{od}(h)$ for path $r \in R^{od}$ during interval $h$ is expressed as follows:

$$\bar{V}_r^{od}(h) = \sum_{a \in A} t_a(h) \cdot \delta_{a,r}^{od} + \frac{\bar{\tau}_r^{od}(h)}{\text{VOT}} \tag{13}$$

where VOT is the average value of time of all users. With the JDTT and JDDT, the generalized travel cost functions $\tilde{V}_r^{od}(h)$ and $\hat{V}_r^{od}(h)$ for path $r \in R^{od}$ during interval $h$ are expressed, respectively, as follows:

$$\tilde{V}_r^{od}(h) = \sum_{a \in A} t_a(h) \cdot \delta_{a,r}^{od} + \frac{\bar{\tau}_r^{od}(h) + \tilde{\tau}_r^{od}(h)}{\text{VOT}} \tag{14}$$

$$\hat{V}_r^{od}(h) = \sum_{a \in A} t_a(h) \cdot \delta_{a,r}^{od} + \frac{\bar{\tau}_r^{od}(h) + \hat{\tau}_r^{od}(h)}{\text{VOT}} \tag{15}$$

When $\beta_1 = \beta_2 = 0$, $\tilde{V}_r^{od}(h) = \hat{V}_r^{od}(h) = \bar{V}_r^{od}(h)$ suggesting that the JDTT and JDDT reduce to the distance toll. When $\alpha = 0$, the JDTT and JDDT turn into a time toll and a delay toll, respectively.

We assume that users adjust their route choice in response to pricing following the C-logit stochastic route choice model. As a variation of the well-known multinomial logit (MNL) model, the C-logit model seeks, by introducing a "commonality factor", to address the concern resulting from the independence from irrelevant alternatives (IIA) property of the MNL model, while maintaining a closed-form expression for the route choice probability (Cascetta, 2001; Cascetta et al., 1996). The route choice probability for path $r \in R^{od}$ during interval $h + 1$ is expressed as follows:

$$p_r^{od}(h+1) = \frac{\exp\left[\theta_0 \cdot \left(-V_r^{od}(h) - CF_r(h)\right)\right]}{\sum_{r \in R^{od}} \exp\left[\theta_0 \cdot \left(-V_r^{od}(h) - CF_r(h)\right)\right]} \tag{16}$$

$$CF_r(h) = \beta_0 \cdot \ln \sum_{\substack{s \in R^{od}, \\ s \neq r}} \left(\frac{L_{rs}(h)}{L_r(h)^{\frac{1}{2}} \cdot L_s(h)^{\frac{1}{2}}}\right)^{\gamma_0} \tag{17}$$

where $V_r^{od}(h)$ is a nominal expression of the generalized travel cost function; $\theta_0$, $\beta_0$, and $\gamma_0$ are positive scale parameters; $CF_r(h)$ is the "commonality factor" for path $r \in R^{od}$ during interval $h$; $L_{rs}(h)$ is the "length" of links common to paths $r$ and $s$ during interval $h$; $L_r(h)$ and $L_s(h)$ are the total "lengths" of paths $r$ and $s$ during interval $h$, respectively ("length" can be either physical or cost-related). The initial route choice probability is determined using the free-flow travel times. Based on the route choice probability, the path assignment is determined as follows:

$$f_r^{od}(h+1) = q^{od}(h+1) \cdot p_r^{od}(h+1)$$
$$= q^{od}(h+1) \cdot \frac{\exp\left[\theta_0 \cdot \left(-V_r^{od}(h) - CF_r(h)\right)\right]}{\sum_{r \in R^{od}} \exp\left[\theta_0 \cdot \left(-V_r^{od}(h) - CF_r(h)\right)\right]} \quad (18)$$

*2.3. Feedback control strategies*

The PI controller is a widely used feedback control strategy for traffic control and management (Keyvan-Ekbatani et al., 2012; Papageorgiou et al., 1991; Yin and Lou, 2009; Zheng et al., 2016; Zheng et al., 2012). It has already been applied to road user pricing but in two different directions: (i) applying the PI controller to real-time (i.e. within-day) pricing, e.g. Yin and Lou (2009); and (ii) applying the PI controller to day-to-day pricing, e.g. Zheng et al. (2016). The biggest differences between the two are:

i. Real-time pricing results in continuously changing toll rates (with a time step applied) due to its real-time nature, whereas day-to-day pricing produces different static toll rates for different tolling intervals.
ii. Real-time pricing does not need an iterative solution framework because for a certain time interval, the inputs to the PI controller come from the previous time interval. Differently, day-to-day pricing needs an iterative solution framework (as will be shown in Section 2.4). For a certain time interval, the inputs to the PI controller always come from the same time interval but in the previous iteration.

Since we use the NFD as an indicator of the network performance, the PI controller becomes a suitable means because the critical network density obtained from the NFD naturally becomes the set point in the PI controller. Therefore, applying the PI controller can effectively achieve this particular objective and solve the optimization problem. However, when the objective changes to, for example, minimizing the total travel time, the PI controller can hardly be applied due to the lack of a set point. We may have to use other numerical optimization methods, e.g. surrogate modeling (Chen et al., 2014).

*2.3.1. Pricing controller for the distance toll*

Under the distance toll scenario, the distance toll rate $\alpha$ is the only control input. To achieve the control objective, $\alpha$ is adjusted iteratively through a discrete PI controller which is mathematically expressed as follows (see Appendix A for derivation):

$$\alpha(i) = \alpha(i-1) + P_p^\alpha \cdot \left(K_{\max}(i) - K_{\max}(i-1)\right) + P_i^\alpha \cdot (K_{\max}(i) - K_{cr}) \quad (19)$$

where $i \in (2, \dots, N_{\max})$ is the number of iteration upper-bounded by the maximum value $N_{\max}$; $K_{\max}(i)$ is the maximum network density within the tolling period of iteration $i$; $P_p^\alpha$ and $P_i^\alpha$ are positive proportional and integral gain parameters for $\alpha$, respectively; $K_{cr}$ is the critical network density identified from the NFD. We use the maximum value rather than the average to be less conservative in pricing. The initial distance toll rate when $i = 1$ is determined as follows:

$$\alpha(1) = P_i^\alpha \cdot (K_{\max}(1) - K_{cr}) \quad (20)$$

where $K_{\max}(1)$ is the maximum network density within the tolling period without pricing. Eq. (19) shows that (i) the pricing controller can be less conservative by replacing $K_{\max}(i)$ with the average network density within the tolling period; and (ii) $\alpha(i)$ is calculated based on the outputs of iterations $i$ and $i-1$, but is implemented in iteration $i+1$.

The discrete PI controller has the following four important properties (Kosmatopoulos and Papageorgiou, 2003; Papageorgiou and Kotsialos, 2002; Zheng and Geroliminis, 2016; Zheng et al., 2016; Zheng et al., 2012):

i. Dynamic traffic networks with small errors in traffic states can be stabilized through this feedback control to the optimal state $K_{\mathrm{cr}}$.
ii. Users' adaptation to pricing is considered. For example, if $K_{\max}(i) < K_{\max}(i-1)$, $P_{\mathrm{p}}^{\alpha} \cdot (K_{\max}(i) - K_{\max}(i-1))$ becomes negative to prevent users from being overcharged.
iii. The gain parameters $P_{\mathrm{p}}^{\alpha}$ and $P_{\mathrm{i}}^{\alpha}$ are determined through trial-and-error. In general, larger values lead to faster convergence but with aggravated oscillations. This implies a trade-off between computational time and convergence smoothness.
iv. Fast and global convergence of this feedback control ensures its robustness to moderate parameter changes and preserves its closed-loop stability.

Here, "convergence" is only valid for the iterative PI controller investigated in this paper as well as in Zheng et al. (2016); Zheng et al. (2012), but not for the real-time PI controller applied to perimeter control (Aboudolas and Geroliminis, 2013; Haddad and Shraiber, 2014; Keyvan-Ekbatani et al., 2012). This is because in perimeter control, the control inputs for managing transfer flows between regions always vary as a result of the changing real-time measurements.

*2.3.2. Pricing controller for the JDTT: a simultaneous approach*

Unlike the distance toll involving a single control input only, the JDTT has dual control inputs $\alpha$ and $\beta_1$. Therefore, we extend the discrete PI controller to a matrix form:

$$\begin{bmatrix} \alpha(i) \\ \beta_1(i) \end{bmatrix} = \begin{bmatrix} \alpha(i-1) \\ \beta_1(i-1) \end{bmatrix} + \begin{bmatrix} P_{\mathrm{p}}^{\alpha} & P_{\mathrm{i}}^{\alpha} \\ P_{\mathrm{p}}^{\beta_1} & P_{\mathrm{i}}^{\beta_1} \end{bmatrix} \begin{bmatrix} K_{\max}(i) - K_{\max}(i-1) \\ K_{\max}(i) - K_{\mathrm{cr}} \end{bmatrix} \qquad (21)$$

where $P_{\mathrm{p}}^{\beta_1}$ and $P_{\mathrm{i}}^{\beta_1}$ are positive proportional and integral gain parameters for $\beta_1$, respectively. Since there are four parameters in Eq. (21), directly applying trial-and-error could lead to different pricing results which, however, are all optimal in the sense that they all help achieve the same control objective (i.e. there could be multiple combinations of the two pairs of parameters). To ensure a unique optimal solution from the PI controller, we need to make a balance between $\alpha$ and $\beta_1$.

Given that the PI controller is robust to moderate parameter changes, we assume that the rank of $\begin{bmatrix} P_{\mathrm{p}}^{\alpha} & P_{\mathrm{i}}^{\alpha} \\ P_{\mathrm{p}}^{\beta_1} & P_{\mathrm{i}}^{\beta_1} \end{bmatrix}$ is equal to 1:

$$\begin{bmatrix} P_{\mathrm{p}}^{\alpha} & P_{\mathrm{i}}^{\alpha} \\ P_{\mathrm{p}}^{\beta_1} & P_{\mathrm{i}}^{\beta_1} \end{bmatrix} = \begin{bmatrix} \mu_{\alpha} & 0 \\ 0 & \mu_{\beta_1} \end{bmatrix} \begin{bmatrix} P_{\mathrm{p}} & P_{\mathrm{i}} \\ P_{\mathrm{p}} & P_{\mathrm{i}} \end{bmatrix} \qquad (22)$$

where $P_{\mathrm{p}}$ and $P_{\mathrm{i}}$ are nominal positive proportional and integral gain parameters, respectively, and $\mu_{\alpha}$ and $\mu_{\beta_1}$ are positive scale parameters to be determined, respectively. Eq. (21) is therefore rewritten as follows:

$$\begin{bmatrix} \alpha(i) \\ \beta_1(i) \end{bmatrix} = \begin{bmatrix} \alpha(i-1) \\ \beta_1(i-1) \end{bmatrix} + \begin{bmatrix} \mu_{\alpha} & 0 \\ 0 & \mu_{\beta_1} \end{bmatrix} \begin{bmatrix} P_{\mathrm{p}} & P_{\mathrm{i}} \\ P_{\mathrm{p}} & P_{\mathrm{i}} \end{bmatrix} \begin{bmatrix} K_{\max}(i) - K_{\max}(i-1) \\ K_{\max}(i) - K_{\mathrm{cr}} \end{bmatrix} \qquad (23)$$

By denoting $\begin{bmatrix} \alpha(i) \\ \beta_1(i) \end{bmatrix}$ as $\boldsymbol{u}(i)$, $\begin{bmatrix} \mu_\alpha & 0 \\ 0 & \mu_{\beta_1} \end{bmatrix}$ as $\boldsymbol{\mu}$, $\begin{bmatrix} P_p & P_i \\ P_p & P_i \end{bmatrix}$ as $\boldsymbol{P}$, and $\begin{bmatrix} K_{\max}(i) - K_{\max}(i-1) \\ K_{\max}(i) - K_{cr} \end{bmatrix}$ as $\begin{bmatrix} E_p(i) \\ E_i(i) \end{bmatrix}$ or $\boldsymbol{E}(i)$, Eq. (23) is further simplified:

$$\boldsymbol{u}(i) = \boldsymbol{u}(i-1) + \boldsymbol{\mu P E}(i) \tag{24}$$

The initial toll rates when $i = 1$ are determined as follows:

$$\boldsymbol{u}(1) = \boldsymbol{\mu P E}(1) = \begin{bmatrix} \mu_\alpha & 0 \\ 0 & \mu_{\beta_1} \end{bmatrix} \begin{bmatrix} P_p & P_i \\ P_p & P_i \end{bmatrix} \begin{bmatrix} 0 \\ E_i(1) \end{bmatrix} = \begin{bmatrix} \mu_\alpha \cdot P_i \cdot E_i(1) \\ \mu_{\beta_1} \cdot P_i \cdot E_i(1) \end{bmatrix} \tag{25}$$

Based on Eq. (24), we propose the following simultaneous approach to iteratively adjust the JDTT.

**Proposition 1.** The ratio between the optimal steady-state toll rates is equal to the ratio between the scale parameters, i.e. $\frac{\alpha(i^*)}{\beta_1(i^*)} = \frac{\mu_\alpha}{\mu_{\beta_1}}$ where $i^*$ is the number of iteration in which the steady-state error comes close to zero.

The proof of Proposition 1 is provided in Appendix B. To implement Eq. (24), we need to specify the relative "weight" between the two toll components. Thanks to the link-additive property, we can rescale our analysis from the path level to the link level. The generalized travel cost function $c_a(h)$ for link $a \in \bar{A}$ during interval $h$ is expressed as follows

$$c_a(h) = t_a(h) + \bar{\tau}_a(h) + \tilde{\tau}_a(h) \tag{26}$$

where $\bar{\tau}_a(h) = \alpha \cdot l_a$ and $\tilde{\tau}_a(h) = \beta_1 \cdot t_a(h)$ are the distance and time toll components for link $a \in \bar{A}$ during interval $h$, respectively. Assuming that the relative "weight" between the optimal steady-state distance and time toll components for link $a \in \bar{A}$ during interval $h$ is $\omega_1$, i.e. $\frac{\bar{\tau}_a^*(h)}{\tilde{\tau}_a^*(h)} = \omega_1$, the following equality holds:

$$\frac{\alpha(i^*)}{\beta_1(i^*)} = \frac{\mu_\alpha}{\mu_{\beta_1}} = \frac{\omega_1 \cdot t_a(h)}{l_a} \tag{27}$$

As a result, the assumption does not hold anymore because the resulting ratio between the optimal steady-state distance and time toll rates is not unique which violates the global convergence law. A remedy instead of looking at an individual link is to aggregate $\frac{\bar{\tau}_a^*(h)}{\tilde{\tau}_a^*(h)}$ over all links and intervals and assume the average to be $\omega_1$:

$$\frac{\sum_{h=j}^{j+N_t-1} \sum_{a \in \bar{A}} \frac{\bar{\tau}_a^*(h)}{\tilde{\tau}_a^*(h)}}{N_t \cdot \bar{n}} = \omega_1 \tag{28}$$

where $j$ is the first tolling interval; $N_t$ is the total number of intervals within the tolling period; $\bar{n}$ is the total number of links in sub-network $\bar{G}$. The remedied assumption can be seen as a network-level extension to the previous one for an individual link. The physical meaning of $\omega_1$ is that it specifies the relative amount of cost incurred by the distance and time toll components, respectively. For example, $\omega_1 = 1$ suggests that the distance and time toll components equally add to the generalized travel cost.

Given Eq. (29):

$$\frac{\bar{\tau}_a^*(h)}{\tilde{\tau}_a^*(h)} = \frac{\alpha(i^*) \cdot l_a}{\beta_1(i^*) \cdot t_a(h)} = \frac{\mu_\alpha \cdot l_a}{\mu_{\beta_1} \cdot t_a(h)} = \frac{\mu_\alpha \cdot v_a(h)}{\mu_{\beta_1}} \tag{29}$$

where $v_a(h)$ is the average speed of link $a \in \bar{A}$ during interval $h$ when $\alpha(i^*)$ and $\beta_1(i^*)$ are imposed, Eq. (28) is rewritten as follows:

$$\frac{\mu_\alpha}{\mu_{\beta_1}} = \frac{\omega_1 \cdot N_t \cdot \bar{n}}{\sum_{h=j}^{j+N_t-1} \sum_{a \in \bar{A}} v_a(h)} = \frac{\omega_1 \cdot N_t}{\sum_{h=j}^{j+N_t-1} \bar{v}(h)} = \frac{\omega_1}{\bar{\bar{v}}} \tag{30}$$

where $\bar{v}(h) = \frac{1}{\bar{n}} \cdot \sum_{a \in \bar{A}} v_a(h)$ is the average speed of the cordon area during interval $h$ when $\alpha(i^*)$ and $\beta_1(i^*)$ are imposed, and $\bar{\bar{v}} = \frac{1}{N_t} \cdot \sum_{h=j}^{j+N_t-1} \bar{v}(h)$ is the average of $\bar{v}(h)$ over all intervals within the tolling period. Since $P_p$ and $P_i$ are adjustable, we use $\mu_\alpha = 1$ which results in $\mu_{\beta_1} = \frac{\bar{\bar{v}}}{\omega_1}$. Although $\bar{\bar{v}}$ is not computable because prior knowledge of $\alpha(i^*)$ and $\beta_1(i^*)$ is required, different tolls may result in similar network speeds observed from the speed-density NFDs. This is because the control logic is always to keep the maximum network density within the tolling period around the critical threshold identified from the NFD. We therefore approximate $\bar{\bar{v}}$ using the simulation outputs of the optimal cordon toll scenario. Validity of this approximation will be shown in Section 3.2. Without loss of generality, we use $\omega_1 = 1$ to optimize the JDTT. A sensitivity analysis on $\omega_1$ will be provided in Section 3.2.1.

### 2.3.3. Pricing controller for the JDDT: a sequential approach

Although the JDDT also has two control inputs $\alpha$ and $\beta_2$, we cannot readily apply the proposed simultaneous approach because $t_a(h)$ in Eq. (11) changes to $t_a(h) - \bar{t}_a$ in Eq. (12). As a result, $v_a(h)$ in Eq. (30) no longer exists suggesting that the NFD-enabled approximation no longer holds. Therefore, we propose the following sequential approach based on Eq. (19) to iteratively adjust the JDDT.

Similar to the introduction of $\omega_1$, we introduce another weight coefficient $\omega_2$ to represent the relative amount of cost incurred by the distance toll component. When $\omega_2 = 1$, $\beta_2 = 0$ and the JDDT reduces to the distance toll. When $\omega_2 = 0$, $\alpha = 0$ and the JDDT turns into a delay toll. Introducing $\omega_2$ helps break down the simultaneous toll level problem to be solved in a sequential manner. We first keep $\beta_2 = 0$ and determine the optimal steady-state distance toll rate $\alpha(i^*)$ through Eqs. (19) and (20). Given $\alpha(i^*)$ and $\omega_2$, we keep $\alpha = \omega_2 \cdot \alpha(i^*)$ and obtain the optimal steady-state delay toll rate $\beta_2(i^*)$ in a similar fashion through Eqs. (31) and (32):

$$\beta_2(i) = \beta_2(i-1) + P_p^{\beta_2} \cdot \left(K_{\max}(i) - K_{\max}(i-1)\right) + P_i^{\beta_2} \cdot \left(K_{\max}(i) - K_{cr}\right) \tag{31}$$

$$\beta_2(1) = P_i^{\beta_2} \cdot \left(K_{\max}(1) - K_{cr}\right) \tag{32}$$

where $P_p^{\beta_2}$ and $P_i^{\beta_2}$ are positive proportional and integral gain parameters for $\beta_2$, respectively. Without loss of generality, we use $\omega_2 = 0.5$ to optimize the JDDT. A sensitivity analysis on $\omega_2$ will be provided in Section 3.3.1.

### 2.4. Simulation-based optimization (SBO) framework

Consider the following equivalent mathematical formulation of the NFD-based toll level problem:

$$\min_{\alpha,\beta} (K_{\max} - K_{cr}) \cdot \text{sgn}(K_{\max} - K_{cr}) \tag{33}$$

subject to

$$0 < \alpha \leq \bar{\alpha} \tag{34}$$

$$0 < \beta \leq \bar{\beta} \tag{35}$$

$$K_{\max} = \text{DTA}(\alpha, \beta) \tag{36}$$

where $\text{sgn}(\cdot)$ is the sign function and $\beta$ is a nominal expression of $\beta_1$ and $\beta_2$. The upper-level objective function in Eq. (33) aims to keep the maximum network density within the tolling period around the critical threshold identified from the NFD. Constraints (34) and (35) specify the feasible regions for adjusting the toll rates where $\bar{\alpha}$ and $\bar{\beta}$ are the upper bounds predefined by a transport authority[4]. Constraint (36) represents the lower-level simulation model taking into account the control inputs $\alpha$ and $\beta$. The simulation output $K_{\max}$ is fed back to the optimization problem and the process continues in an iterative manner until a certain termination criterion is satisfied.

In view of the non-convexity, non-linearity, and non-closed form of the objective function, a new class of solution algorithms termed SBO has recently been applied as a computationally efficient method to solve large-scale urban transportation problems (Osorio and Bierlaire, 2013; Osorio and Chong, 2015). In particular, when SBO is applied to solve the optimal toll level problem, it can be classified into two main categories: SBO using surrogate models (Chen et al., 2016; Chen et al., 2014; Ekström et al., 2016; He et al., 2017) and SBO using feedback control strategies (Simoni et al., 2015; Zheng et al., 2016; Zheng et al., 2012). In this paper, we use the latter approach given its faster and better global convergence compared with the former[5].

An illustration of the developed SBO framework is presented in Fig. 1. The static framework is similar to the one proposed in Zheng et al. (2012) which serves as a fundamental basis for solving time-dependent pricing. Here, the plant refers to the simulation model which is described in Appendix C. The detailed algorithmic steps are provided below:

i. Given a single-cordon two-region network, run the simulation without pricing to get the initial NFD of the cordon area.
ii. Set $i = 1$. Identify from the NFD (i) the critical density $K_{\text{cr}}$; (ii) the tolling period when the network density exceeds $K_{\text{cr}}$; and (iii) the maximum network density within the tolling period $K_{\max}(i)$.
iii. Determine the initial toll through (i) Eq. (202020) for the distance toll; (ii) Eq. (25) for the JDTT; or (iii) Eqs. (20) and (32) for the JDDT.
iv. Set $i = i + 1$. Run the simulation with pricing to get the updated NFD of the cordon area and the new $K_{\max}(i)$.
v. If $i < N_{\max}$, determine the updated toll through (i) Eq. (19) for the distance toll; (ii) Eq. (24) for the JDTT; or (iii) Eqs. (19) and (31) for the JDDT, and go back to step iv; otherwise terminate the algorithm.

---

[4] Under the distance toll scenario, constraint (35) is simply $\beta = 0$.
[5] Surrogate models approximate the response surface for the simulation input-output mapping. The global optimal solution is not guaranteed but is infinitely approached by a sub-optimum with an increasing number of sampling data (Chen et al., 2014). The resulting computational intensity remains challenging particularly for large-scale networks.

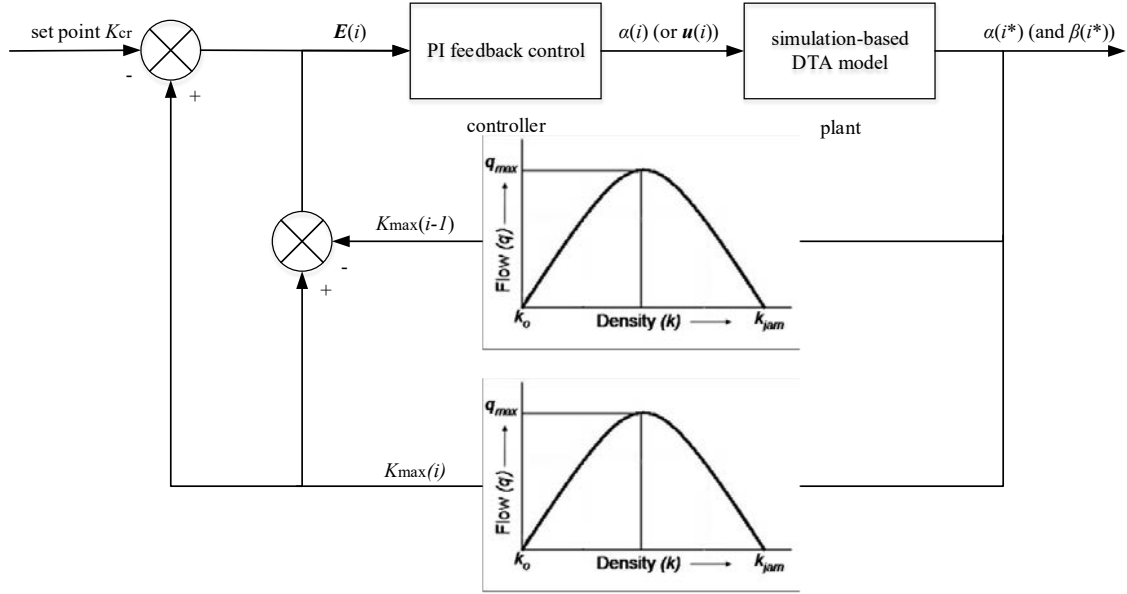

**Fig. 1.** A closed-loop block diagram of the developed SBO framework.

The developed SBO framework can be easily extended to time-dependent pricing which aims to charge drivers according to the changing level of congestion. This is achieved by implementing an "independent" PI controller in each tolling interval. Here, "independent" means that for a specific tolling interval, the input measurements to the PI controller only come from the same interval with no consideration for the other intervals. As a result, there is no coordination among different intervals and an additional label representing the number of interval is to be added to the PI controller. The extended framework is similar to the one proposed in Simoni (2013) and Zheng et al. (2016), but with a significant difference in the pricing regime and hence the way the PI controller is implemented. The biggest advantage of time-dependent pricing is that drivers are not overcharged while the control objective is still met. That is, time-dependent pricing does not over-control the network resulting in unnecessary reduction in flow/production.

If the distribution of congestion within the cordon exhibits strong heterogeneity, further network partitioning (Ji and Geroliminis, 2012; Saeedmanesh and Geroliminis, 2016, 2017) may be needed which results in a coordinated multi-area pricing scheme (Zheng et al., 2016). This is challenging and remains open for future research. In this paper, the feedback control strategies are designed to target the cordon area only with no consideration for the coordination among multiple regions.

## 3. Numerical results and discussion

To quantitatively analyze and compare the performance of different pricing strategies, we perform a case study using a recently developed and calibrated large-scale simulation-based DTA model of Melbourne, Australia. The model is deployed in AIMSUN with time-dependent demand for the 6-10 morning peak (see Appendix C for further details).

When we implement different pricing strategies, users' route choice following the C-logit model is updated every 5 minutes with $\theta_0 = 1$, $\beta_0 = 0.15$, and $\gamma_0 = 1$ (same parameter set-up as in Keyvan-Ekbatani et al. (2015)). We assume VOT = \$15/h (Legaspi and Douglas, 2015). We first run the simulation without pricing to obtain the initial NFD of the cordon area, and to identify the critical network density and the tolling period. We then implement a cordon-based pricing scheme as in Zheng et al. (2016) for comparison with the distance toll. The results and discussion of the non-tolling and optimal cordon toll scenarios are presented in Appendix D.

*3.1. Distance toll*

A limitation of the cordon toll is that the distance traveled within the cordon is not taken as a determinant. Users are equally charged regardless of their actual usage of the urban road space. The resulting social inequity may create negative public acceptability toward the cordon charge. To address this equity concern, we apply the distance-based pricing scheme where the toll price varies by the trip length within the cordon rather than being pay-per-entry. As a result, the distance toll distinguishes, for example, between a user who reaches the destination immediately upon crossing the cordon and a user who traverses the whole cordon area, thereby creating a more efficient and equitable pricing scheme (Daganzo and Lehe, 2015).

Fig. 2 shows the simulation results of the cordon area when the optimal distance toll is implemented. The optimal distance toll rate is around $1.05/km to achieve the control objective. A closer look at Fig. 2(a), however, reveals that the distance toll somehow exacerbates the hysteresis loop in the NFD, the size of which becomes larger than that under the non-tolling scenario. When the network is unloading (i.e. recovering from congestion), the distribution of congestion tends to be more uneven because the congested areas clear more slowly and are fragmented. The uneven distribution of congestion inevitably reduces the network flow during recovery resulting in a clockwise hysteresis loop in the NFD (Gayah and Daganzo, 2011). Given that users choose their routes with the least generalized travel cost, the distance toll naturally drives users into the shortest paths within the cordon because a shorter trip length results in a lower toll price. A large proportion of users therefore travel on the same shortest paths and the distribution of congestion becomes more heterogeneous within the cordon. Fig. 2(d) shows that when compared with the cordon toll, the distance toll results in a much less total distance traveled within the cordon, as expected.

A large hysteresis loop in the NFD is undesirable because the network flow drops significantly during recovery. We show in Fig. 2(e) and (f) the spread-accumulation relationships and time series of the deviation from spread[6] to quantitatively analyze and compare the level of uneven distribution of congestion. Corresponding to the clockwise hysteresis loop in the NFD, an anticlockwise hysteresis loop forms in the spread-accumulation relationship, the size of which increases under the distance toll scenario. The spatial spread of density increases sharply after the imposition of the distance toll and stays at a much higher level. This finding is consistent with Simoni et al. (2015) who argued that the decrease in the network production is generated by clusters of congestion rather than the increase of users.

---

[6] See Appendix E for derivation of $\gamma(k)$ in Eq. (4).

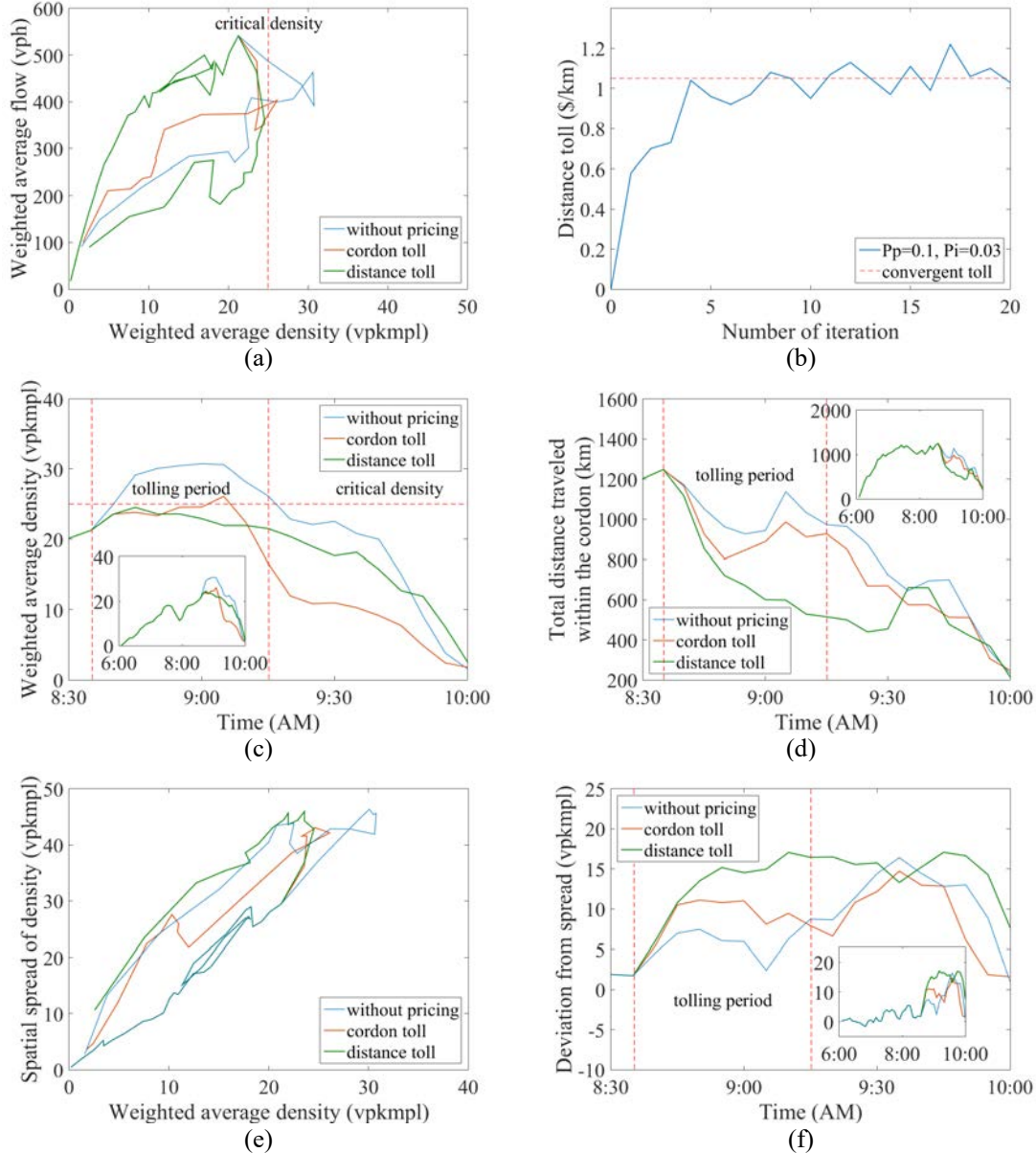

**Fig. 2.** Simulation results of the cordon area when the optimal distance toll is implemented: (a) simulated NFDs; (b) convergence pattern of the distance toll rate; (c) time series of the average network density; (d) time series of the total distance traveled; (e) spread-accumulation relationships; (f) time series of the deviation from spread.

*3.2. Joint distance and time toll (JDTT)*

The question here is how to improve the distance-based pricing scheme so that the network exhibits less heterogeneous distribution of congestion and the resulting NFD has a smaller hysteresis loop. We show that users are driven into the shortest paths within the cordon when charged with the distance toll. Although the travel time on the shortest paths increases as a result of a larger traffic volume, the majority of users do not change their routes because the utility from paying a lower distance toll dominates the disutility from the increase in travel time. The concentration of users into a few shortest paths within the cordon results in a large hysteresis loop in the NFD.

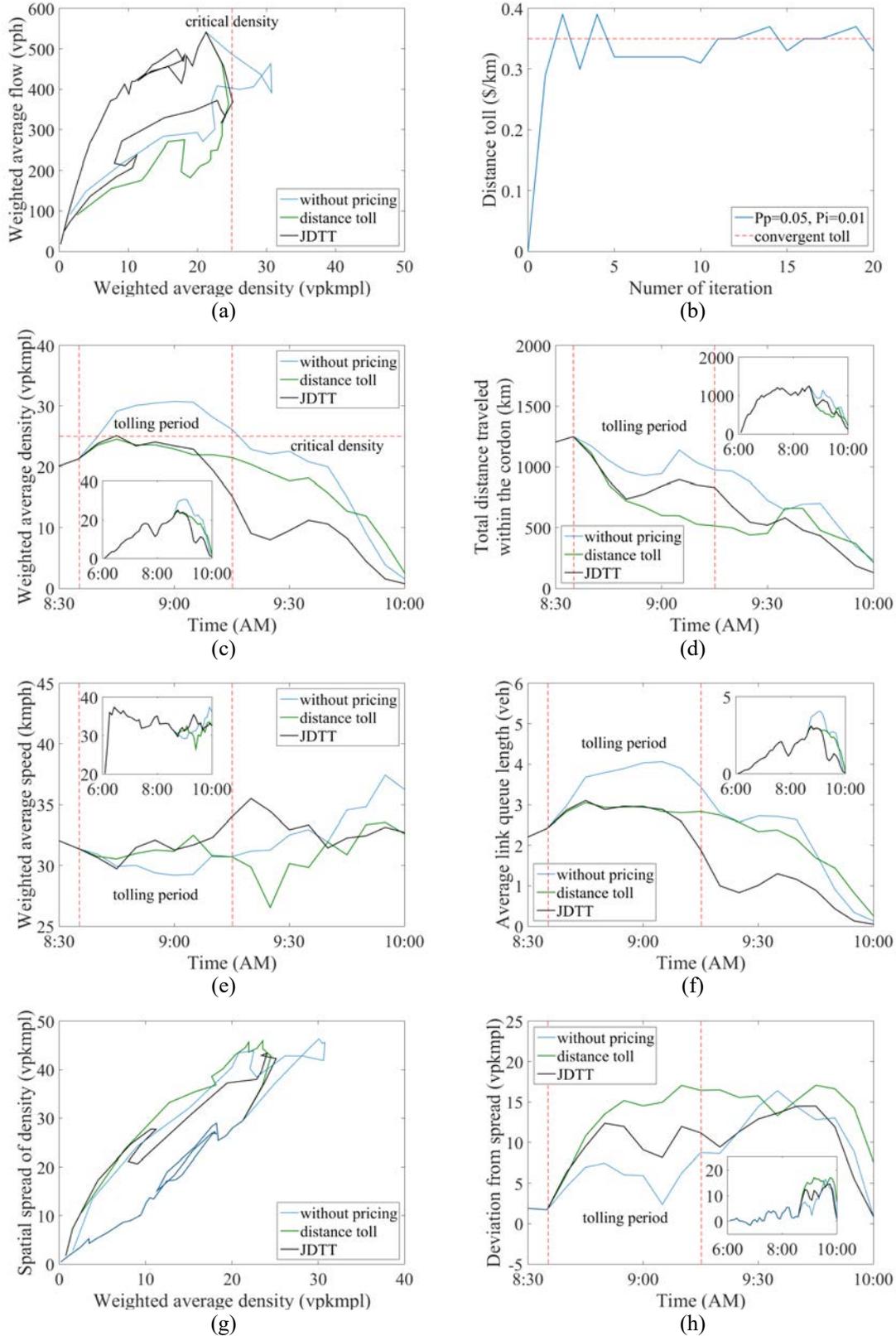

**Fig. 3.** Simulation results of the cordon area when the optimal JDTT is implemented: (a) simulated NFDs; (b) convergence pattern of the distance toll rate; (c) time series of the average network density; (d) time series of the total distance traveled; (e) time series of the average network speed; (f) time series of the average link queue length; (g) spread-accumulation relationships; (h) time series of the deviation from spread.

A solution is to charge users jointly based on the distance traveled and time spent within the cordon. Under the JDTT scenario, users are more likely to distribute themselves into the second or third shortest path if the travel time on the shortest path rises considerably. The uneven distribution of congestion within the cordon may reduce resulting in an NFD with a less distinct hysteresis loop. The JDTT was recently studied in Liu et al. (2014) but only on a static traffic network within the traditional second-best pricing framework. Applying the JDTT to a dynamic traffic network through the proposed NFD-based approach offers a significant contribution.

Fig. 3 shows the simulation results of the cordon area when the optimal JDTT is implemented. To achieve the control objective, the optimal distance toll rate is around 0.35 $/km and the optimal time toll rate is around 9 $/h. Despite having a different order of magnitude, the time toll component exhibits the same convergence pattern as the distance toll component due to their linear correlation. We therefore show the convergence pattern of the distance toll component only. A closer look at Fig. 3(a), (g) and (h) shows that under the JDTT scenario, the size of the hysteresis loop reduces and the deviation from spread stays at a lower level. Compared with the distance toll, the JDTT increases the total distance traveled within the cordon, as expected, because users no longer accumulate themselves into the shortest paths. To further compare the performance of different tolls, we use and show two other evaluation criteria in Fig. 3(e) and (f), respectively: the average network speed and average link queue length. There is a sudden jump in the speed profile at the beginning of the simulation because the network is loading from an empty system without a warm-up period. Although the differences are small immediately after the imposition of different tolls, the time series show that later in time, the JDTT keeps the network speed at a higher level while reducing the link queue length to the greatest extent. The distance toll, however, results in a lower and less stable network speed and an increased link queue length. This network performance degradation results from the fact that clusters of congested links form within the cordon due to the concentration of users into a few specific routes. As shown in Fig. 4, the distance toll results in a more heterogeneous distribution of congestion particularly in the bottom left corner of the cordon area and part of the connected periphery. The differences are not significant at the beginning of the tolling period 8:40 AM, but gradually become prominent later in time toward the end of the tolling period 9:10 AM.

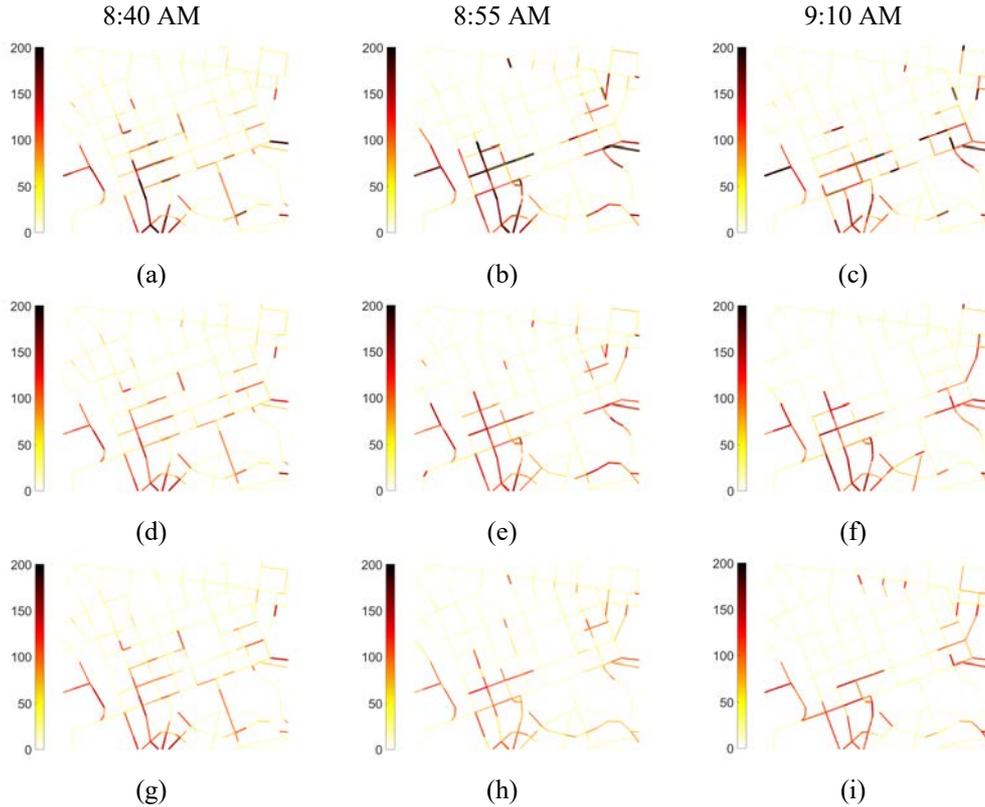

**Fig. 4.** A comparison of the spatiotemporal evolution of link densities within the cordon during the tolling period under different pricing scenarios: (a-c) without pricing; (d-f) distance toll; (g-i) JDTT.

Since we approximate $\bar{\bar{v}}$ using speeds from the optimal cordon toll scenario, we further calculate, for both tolls, the average network speed over the tolling period and end up with $\bar{\bar{v}} = 32.37$ km/h and $\tilde{v} = 32.17$ km/h, thereby justifying the validity of the approximation. When implementing the JDTT, we relax the upper-bound constraints on $\alpha$ and $\beta_1$ assuming that the adjusted toll rates do not exceed the upper limits during the iterations. Nevertheless, if one of the toll rates hits the upper bound during an iteration, it is set fixed at the maximum value during the subsequent iterations and only the other toll rate is to be adjusted. If the control objective is not achieved when both toll rates hit their respective upper bounds, it simply means that pricing alone cannot drive the network to its optimal state and that some complementary measures are needed to create a mixed network control scheme.

### 3.2.1. Sensitivity analysis on $\omega_1$

We perform a sensitivity analysis on $\omega_1$ as a design parameter to examine its effect on the pricing control results. Three different values of $\omega_1$ are tested, i.e. $\omega_1 = 0.33, 1, 3$. A larger value implies a more dominating role played by the distance toll component. Fig. 5(b) and (c) show that regardless of the value of $\omega_1$, the global convergence law consistently holds and that the network is controlled to stay at its optimal state. However, as shown in Fig. 5(a) and (d), both the size of the hysteresis loop and deviation from spread increase when $\omega_1 = 3$. This is because when a large $\omega_1$ is used, the JDTT becomes similar to the distance toll which cannot well reduce the uneven distribution of congestion. Therefore, a small value of $\omega_1$ is advisable for implementing the JDTT. Explicitly optimizing $\omega_1$ is difficult because (i) an analytical mapping is required between $\omega_1$ and an objective function (e.g. minimizing the maximum deviation from spread); and (ii) convexity of the objective function (i.e. solution uniqueness) is not necessarily guaranteed suggesting that multiple local minima may exist. Surrogate modeling is a possible solution but needs further investigation.

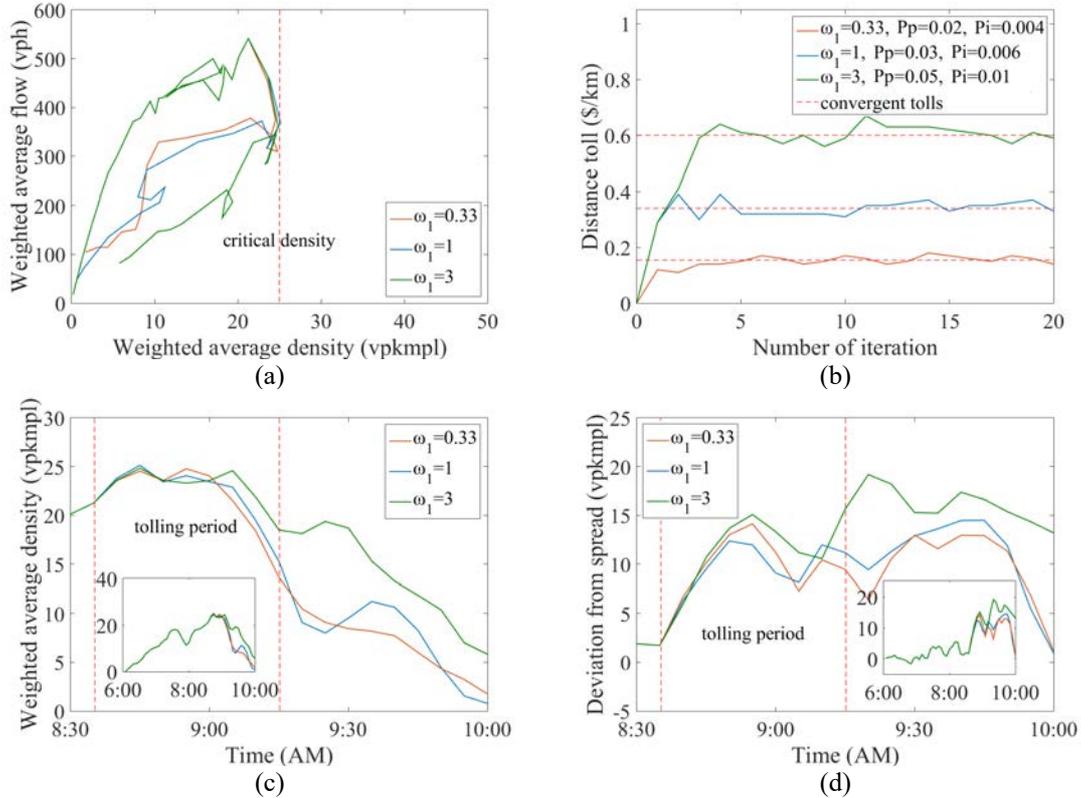

**Fig. 5.** A sensitivity analysis on $\omega_1$: (a) simulated NFDs; (b) convergence patterns of the distance toll rate; (c) time series of the average network density; (d) time series of the deviation from spread.

### 3.2.2. Time-dependent JDTT

While the proposed methodology is first applied to static pricing, we further extend the method under time-dependent pricing where a single tolling interval is partitioned into multiple intervals (two intervals in this study) with a duration of 20 minutes. This duration is not obtained from an optimization problem but is simply selected based on practical implications to demonstrate the dynamic pricing strategy. Intuitively, the duration should not be set too small because drivers could hardly adapt to the rapidly changing toll rates and the network might become unstable (i.e. poor user adaptation). The duration should also not be set too large because different levels of congestion could not be well captured and distinguished (i.e. poor congestion management). The current ERP system in Singapore changes the toll rate every half an hour before and after the peak periods (Liu et al., 2013).

As discussed in Section 2.4, an "independent" PI controller is implemented in each tolling interval and the toll rates for different tolling intervals are iteratively updated in a respective manner until the control objective is met. While the optimal time-dependent JDTT effectively achieves the control objective, Fig. 6(b) shows an interesting non-smooth convergence pattern for the second tolling interval. This is because during the first few iterations, the toll rates in both intervals naturally increase to prevent users from entering the cordon. Due to the interplay between the two intervals, an increase in the toll price for the first interval inevitably leads to a less congested network for the second interval. When the toll price for the first interval comes close to convergence, the toll price for the second interval drops, as expected.

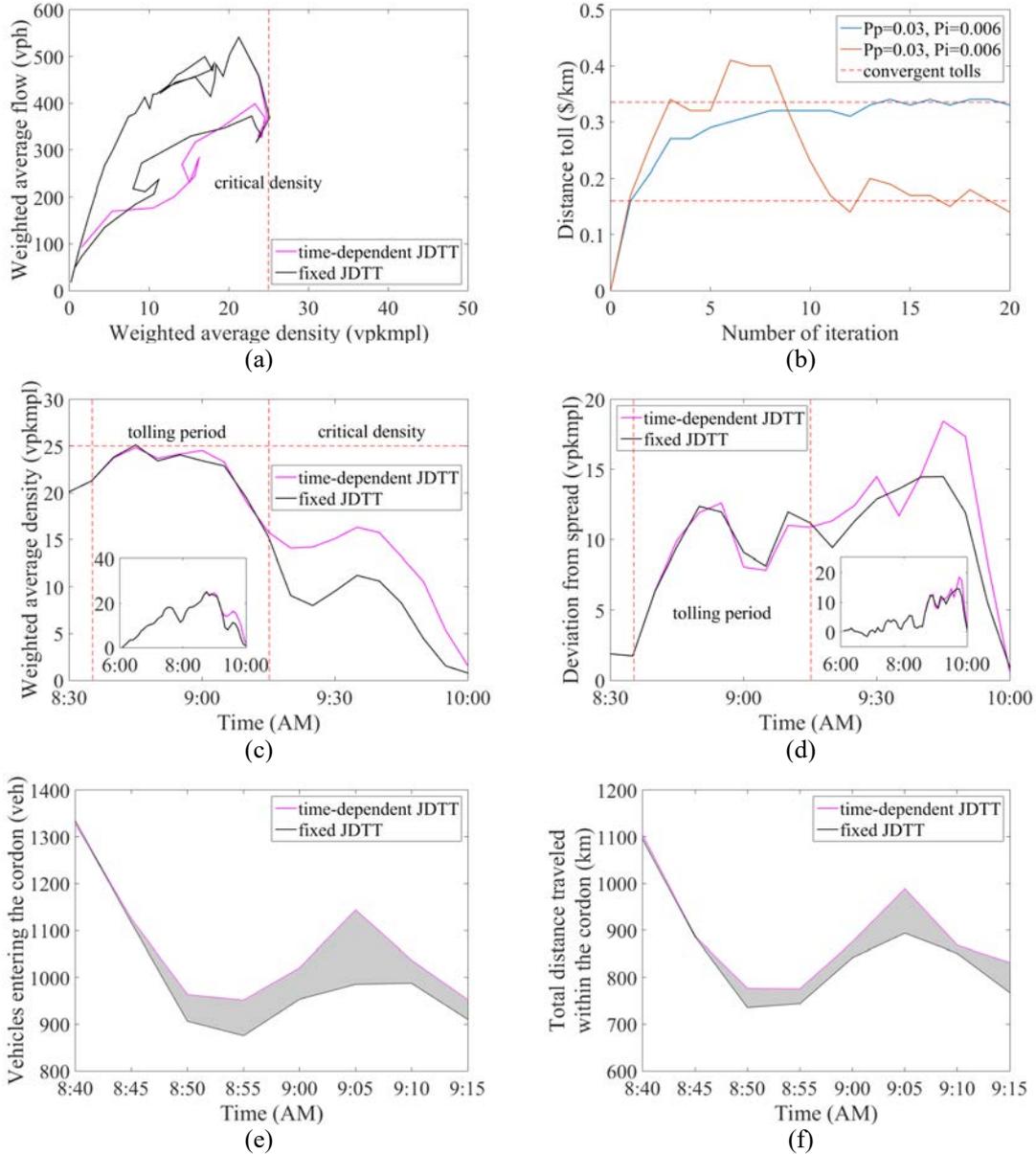

**Fig. 6.** Simulation results of the cordon area when the optimal time-dependent JDTT is implemented: (a) simulated NFDs; (b) convergence patterns of the distance toll rate; (c) time series of the average network density; (d) time series of the deviation from spread; (e) time series of the number of entering vehicles during the tolling period; (f) time series of the total distance traveled during the tolling period.

The advantage of time-dependent pricing is that users are not overcharged because the toll price varies according to the changing level of congestion. Compared with the fixed JDTT, the time-dependent JDTT is less conservative allowing more users to enter the cordon while still achieving the control objective. This is reflected in Fig. 6(e) where the time-dependent JDTT results in a larger number of vehicles entering the cordon during the tolling period, and particularly during the second interval (see the shaded area). Fig. 6(f) compares the total distance traveled within the cordon during the tolling period. As expected, the time-dependent JDTT increases the road usage in the cordon compared with the fixed JDTT. The time-dependency nature improves the equity and hence acceptability of the pricing scheme. The advantage is more significant if the tolling period is long enough to capture different levels of congestion.

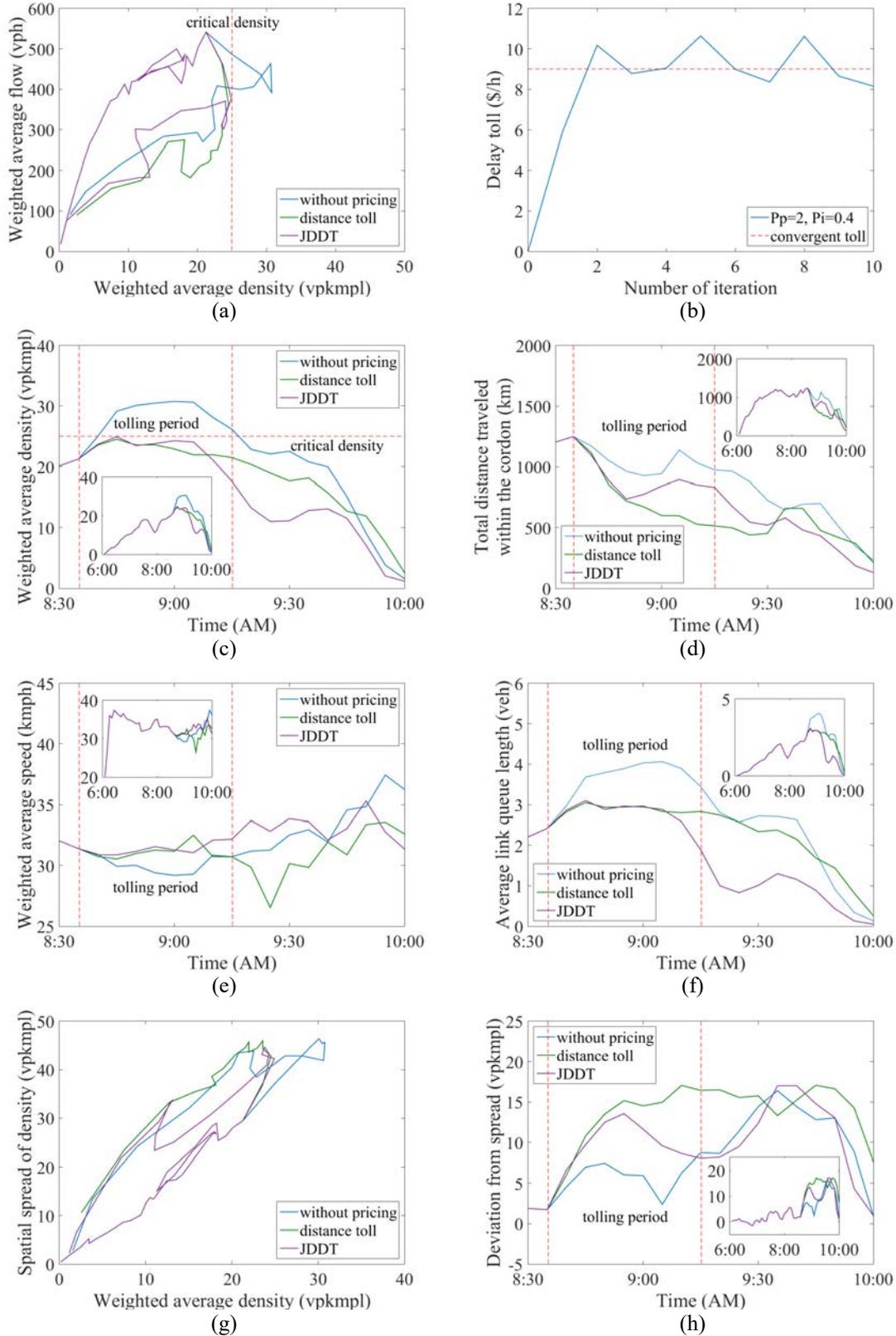

**Fig. 7.** Simulation results of the cordon area when the optimal JDDT is implemented: (a) simulated NFDs; (b) convergence patterns of the toll rates; (c) time series of the average network density; (d) time series of the total distance traveled; (e) time series of the average network speed; (f) time series of the average link queue length; (g) spread-accumulation relationships; (h) time series of the deviation from spread.

## 3.3. Joint distance and delay toll (JDDT)

A limitation of the JDTT is that the time toll component tends to overcharge users because a longer link typically requires more travel time despite being uncongested. Charging users partially in proportion to their travel delays is therefore more sensible. The resulting JDDT is a further improved pricing model that also addresses the limitation of the distance toll. Fig. 7 shows the simulation results of the cordon area when the optimal JDDT is implemented. The optimal distance toll rate is around 0.5 $/km and the optimal delay toll rate is around 9 $/h. While achieving the control objective, the JDDT reduces the uneven distribution of congestion within the cordon resulting in a less distinct hysteresis loop in the NFD, as expected. This is reflected quantitatively in Fig. 7(g) and (h) and qualitatively in Fig. 8. Results so far seem to suggest that the JDTT and JDDT perform equally well in controlling the network. We therefore perform a further comparison between the JDTT and JDDT in Section 3.3.2.

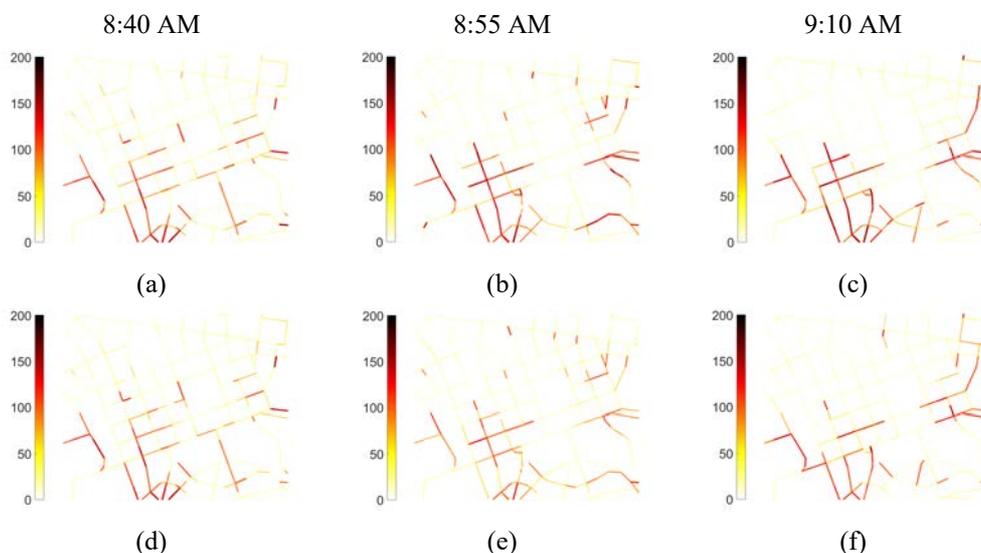

**Fig. 8.** A comparison of the spatiotemporal evolution of link densities within the cordon during the tolling period between (a-c) the distance toll and (d-f) JDDT.

### 3.3.1. Sensitivity analysis on $\omega_2$

Similar to Section 3.2.1, we perform a sensitivity analysis on $\omega_2$ as a design parameter to examine its effect on the pricing control results. Three different values of $\omega_2$ are tested, i.e. $\omega_2 = 0.25, 0.5, 0.75$. A larger value implies a more dominating role played by the distance toll component. Fig. 9 shows the simulation results using different values of $\omega_2$. Since the convergence pattern of the distance toll component remains the same regardless of the value of $\omega_2$, we only present the convergence patterns of the delay toll component. The network performance does not vary significantly as $\omega_2$ changes because the simulated NFDs and time series exhibit similar patterns. Therefore, the pricing control results are not sensitive to different values of $\omega_2$.

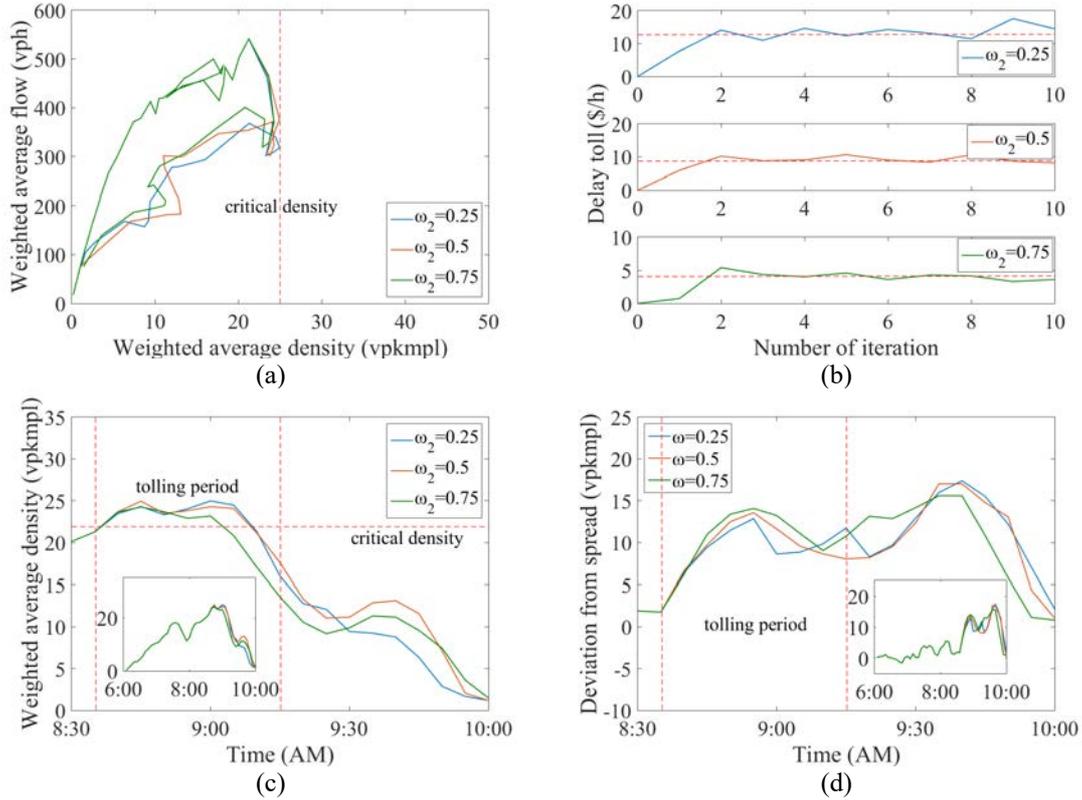

**Fig. 9.** A sensitivity analysis on $\omega_2$: (a) simulated NFDs; (b) convergence patterns of the delay toll rate; (c) time series of the average network density; (d) time series of the deviation from spread.

### 3.3.2. Comparison between the JDTT and JDDT

Since the JDTT and JDDT are implemented through the simultaneous approach and the sequential approach, respectively, we further apply the sequential approach to the JDTT and compare the following three different pricing strategies: (i) JDTT using the simultaneous approach; (ii) JDTT using the sequential approach; and (iii) JDDT using the sequential approach.

Fig. 10 shows that the network performance exhibits similar patterns under the three pricing strategies. Both the JDTT and JDDT reduce the uneven distribution of congestion within the cordon while achieving the control objective. Nevertheless, since the JDDT does not overcharge users on longer links in the network (which typically require more travel time but does not necessarily mean more travel delay), it is supposed to outperform the JDTT resulting in a better network performance. A possible reason why the difference is not significant may be because very few links within the cordon have an extra-long length. The difference may be more significant in a network where the link length varies considerably (i.e. having a greater link length heterogeneity), but this needs further investigation.

Table 2 shows a few selected network performance measures under different pricing scenarios. When considering the entire network, the studied pricing models result in similar average travel times and speeds which are almost the same as those under the non-tolling scenario. The reason why the overall network performance is not improved much is because the predefined cordon only covers a relatively small area of the entire network (i.e. the congested city center of Melbourne) and hence, the effect of pricing is not significant by simply referring to those performance measurements of the entire network. With an increasing size of the cordon, the effect of pricing on the entire network shall become more significant. Nevertheless, although the difference in the average travel time is not significant, we could calculate the total travel time saving by multiplying it by the total number of vehicles which gives us more than 1,000 hours of travel time saving as compared with the 4-hour morning peak. When we look at the cordon area rather than the entire network, the difference in the average travel time becomes larger, as expected. Since the average travel time in the cordon is in the range of 3-4 minutes, an 11-14% average travel time saving is achieved for all the pricing models except the distance toll. The distance toll improves the average travel time in the cordon by 6% only which is equivalent to a 5-10% increase in the average travel time in the cordon compared with the other tolls. This observation is consistent with the fact that the average travel speeds in the cordon under all the other pricing scenarios are improved by 5-14% compared with that under the non-tolling scenario, whereas the distance toll reduces the average travel speed in the cordon. Also, as expected, the distance toll generates the lowest average distance traveled in the cordon.

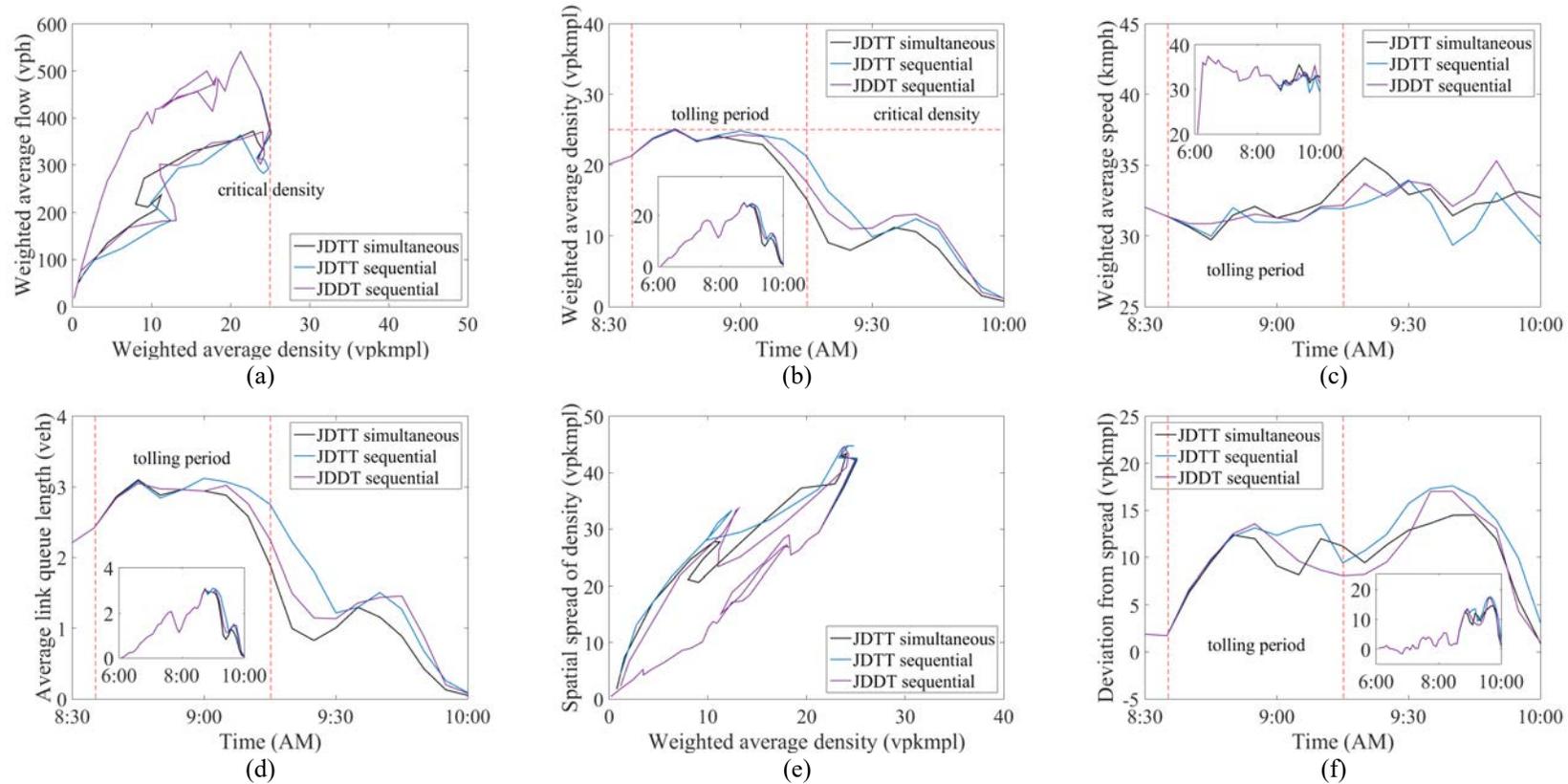

**Fig. 10.** A comparison between the JDTT and JDDT: (a) simulated NFDs; (b) time series of the average network density; (c) time series of the average network speed; (d) time series of the average link queue length; (e) spread-accumulation relationships; (f) time series of the deviation from spread.

Table 2 Selected network performance measures under different pricing scenarios.

| Network performance measures | | No pricing | Cordon toll | Distance toll | Time toll | Delay toll | JDTT (simultaneous) | JDTT (sequential) | JDDT (sequential) |
|---|---|---|---|---|---|---|---|---|---|
| Total number of vehicles simulated (veh) | Entire network | 348,789 | 348,211 | 346,671 | 348,455 | 347,524 | 348,546 | 348,201 | 348,082 |
| | Cordon area | 53,749 | 53,096 | 52,673 | 53,145 | 53,232 | 53,121 | 52,633 | 53,549 |
| Total travel time (h) | Entire network | 94,295 | 94,290 | 92,383 | 93,370 | 93,829 | 95,024 | 94,788 | 93,784 |
| | Cordon area | 3,763 | 3,189 | 3,464 | 3,200 | 3,243 | 3,216 | 3,286 | 3,245 |
| Total distance traveled (km) | Entire network | 2,169,537 | 2,182,403 | 2,146,487 | 2,160,127 | 2,162,866 | 2,185,630 | 2,175,822 | 2,165,691 |
| | Cordon area | 42,909 | 41,382 | 38,814 | 40,240 | 40,691 | 39,930 | 39,438 | 40,226 |
| Average distance traveled (km/veh) | Entire network | 6.22 | 6.27 | 6.19 | 6.20 | 6.22 | 6.27 | 6.25 | 6.22 |
| | Cordon area | 0.80 | 0.78 | 0.74 | 0.76 | 0.76 | 0.75 | 0.75 | 0.75 |
| Average travel time (min/veh) | Entire network | 16.22 | 16.25 | 15.99 | 16.08 | 16.20 | 16.36 | 16.33 | 16.17 |
| | Cordon area | 4.20 | 3.60 | 3.95 | 3.61 | 3.66 | 3.63 | 3.75 | 3.64 |
| Average travel speed (km/h) | Entire network | 23.01 | 23.15 | 23.23 | 23.14 | 23.05 | 23.00 | 22.95 | 23.09 |
| | Cordon area | 11.40 | 12.98 | 11.20 | 12.58 | 12.55 | 12.42 | 12.00 | 12.40 |

*3.4. Simulation stochasticity*

Results so far suggest that the studied pricing models can effectively achieve the control objective. The resulting network performance, however, shows a major difference in the size of the hysteresis loop in the NFD. Since the hysteresis loop is an effect of uneven distribution of congestion, the deviation from spread is a key criterion for quantitatively evaluating different pricing models.

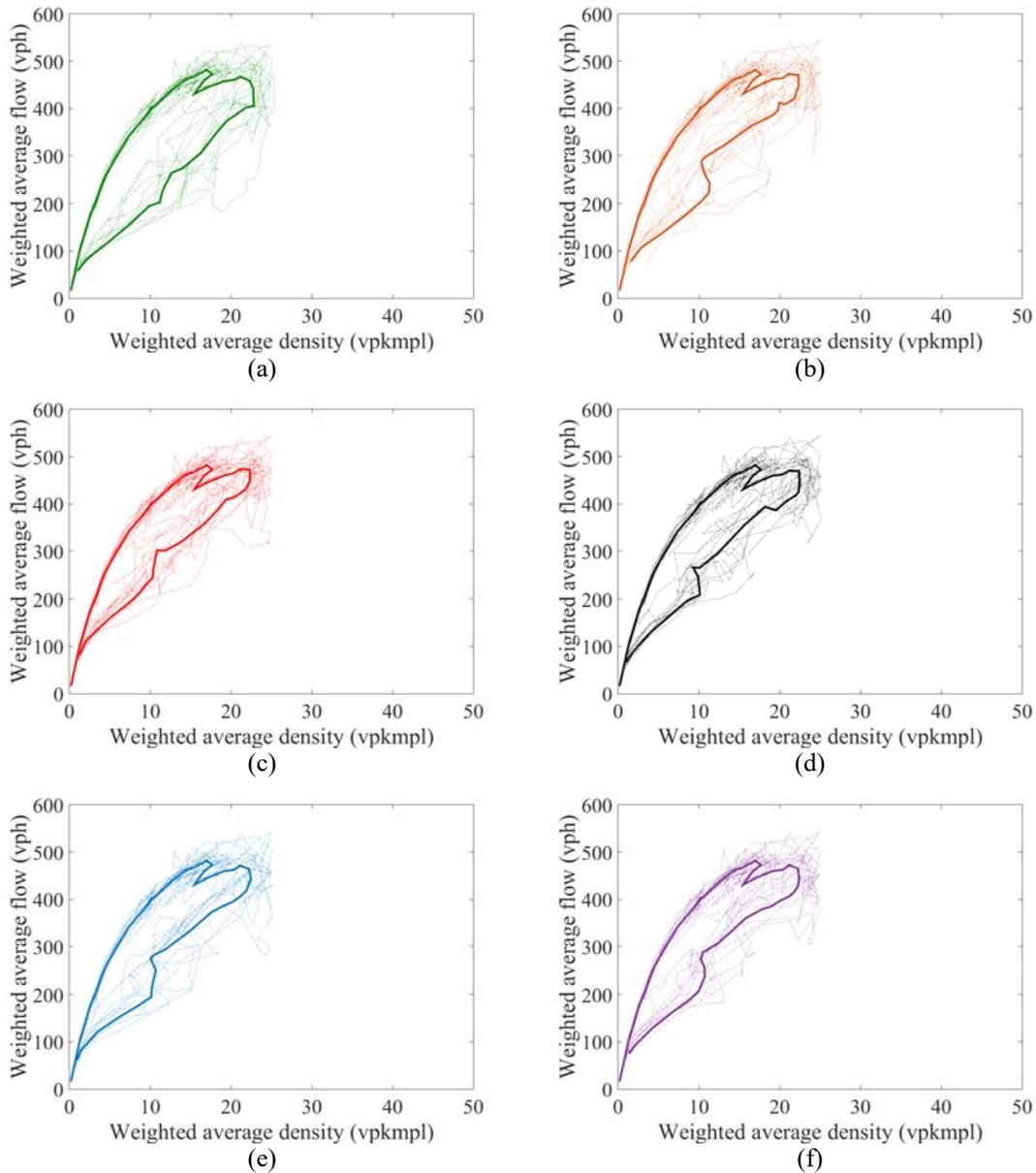

**Fig. 11.** Averaged simulated NFDs with ten different random seed numbers under different pricing scenarios: (a) distance toll; (b) time toll; (c) delay toll; (d) JDTT simultaneous; (e) JDTT sequential; (f) JDDT sequential.

To account for the stochastic effect of simulation, we further apply the developed SBO framework with ten different random seed numbers under different pricing scenarios. Fig. 11 shows that all the studied pricing models successfully keep the network from entering the congested regime of the NFD. A closer look at Fig. 12(a) reveals that all the other pricing models outperform the distance toll because of the less distinct hysteresis loop. This is also reflected in Fig. 12(b) which shows the distributions of the maximum deviation from spread under different pricing scenarios. As expected, the distance toll generates the highest deviation from spread while all the other pricing models perform similar. From a control perspective, all the studied pricing models are effective in reducing congestion within the cordon against simulation stochasticity. However, from a traffic engineering perspective, the JDTT and JDDT are more desirable than the distance toll due to their capability of reducing the uneven distribution of congestion and hence of better maintaining the network stability. While both the time toll and delay toll perform equally well as the JDTT and JDDT, we do not recommend implementing such a time- or delay-based only pricing scheme, because it may result in safety and environmental concerns by encouraging users to drive more aggressively and to use minor roads (May and Milne, 2000).

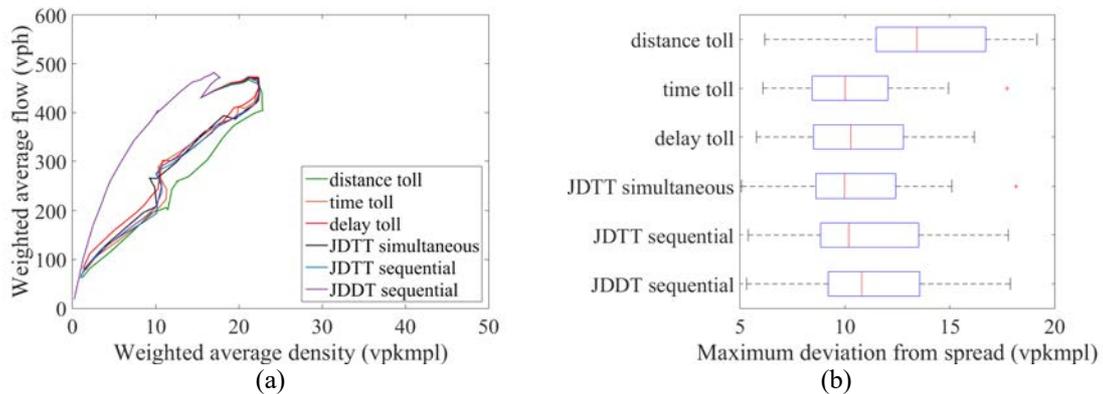

**Fig. 12.** A comparison of the simulation results of the cordon area under different pricing scenarios: (a) averaged simulated NFDs; (b) a box plot showing the distributions of the maximum deviation from spread during the tolling period.

## 4. Applicability of the feedback control: global convergence guaranteed?

Throughout the analysis, we have repeatedly seen the global convergence of the feedback control. The question is whether the global convergence holds under all traffic scenarios. The answer is no. The feedback control is applicable with guaranteed convergence only if a prerequisite is satisfied: the periphery of the cordon area should have sufficient capacity to accommodate the re-routed traffic. If the periphery becomes highly congested or gridlocked, it is likely that the pricing control fails.

To demonstrate the prerequisite, we use the cordon toll as an example and repeatedly apply the developed SBO framework with incrementally increasing demand (from 100% to 135% to manually create unreal congestion). When demand increases, the network becomes more congested and the peak-spreading phenomenon becomes more significant. Accordingly, the toll price increases and the tolling period extends. As shown in Fig. 13, the pricing control is able to keep the network from entering the congested regime of the NFD even when demand is relatively high. However, attention should be paid to Fig. 13(d) where a network reloading process first appears in the cordon area reflected by the shape of the NFD. With a further increase in demand, the network reloading process becomes more prominent and the periphery of the cordon comes closer to gridlock. As a result, users are driven back into the cordon, although they have to pay. Under this particular traffic scenario, the application of the feedback control worsens the traffic conditions outside the cordon. The highly congested periphery forces users to re-enter the cordon and the pricing control can no longer price users off the cordon area to achieve the control objective. This is essentially a paradox where the toll price keeps rising but users still enter the cordon. Due to the violation of the prerequisite, the feedback control does not necessarily result in a convergent solution. To show that different tolls do not result in degraded traffic conditions outside the cordon, we further present the simulated NFDs of the periphery in Appendix F.

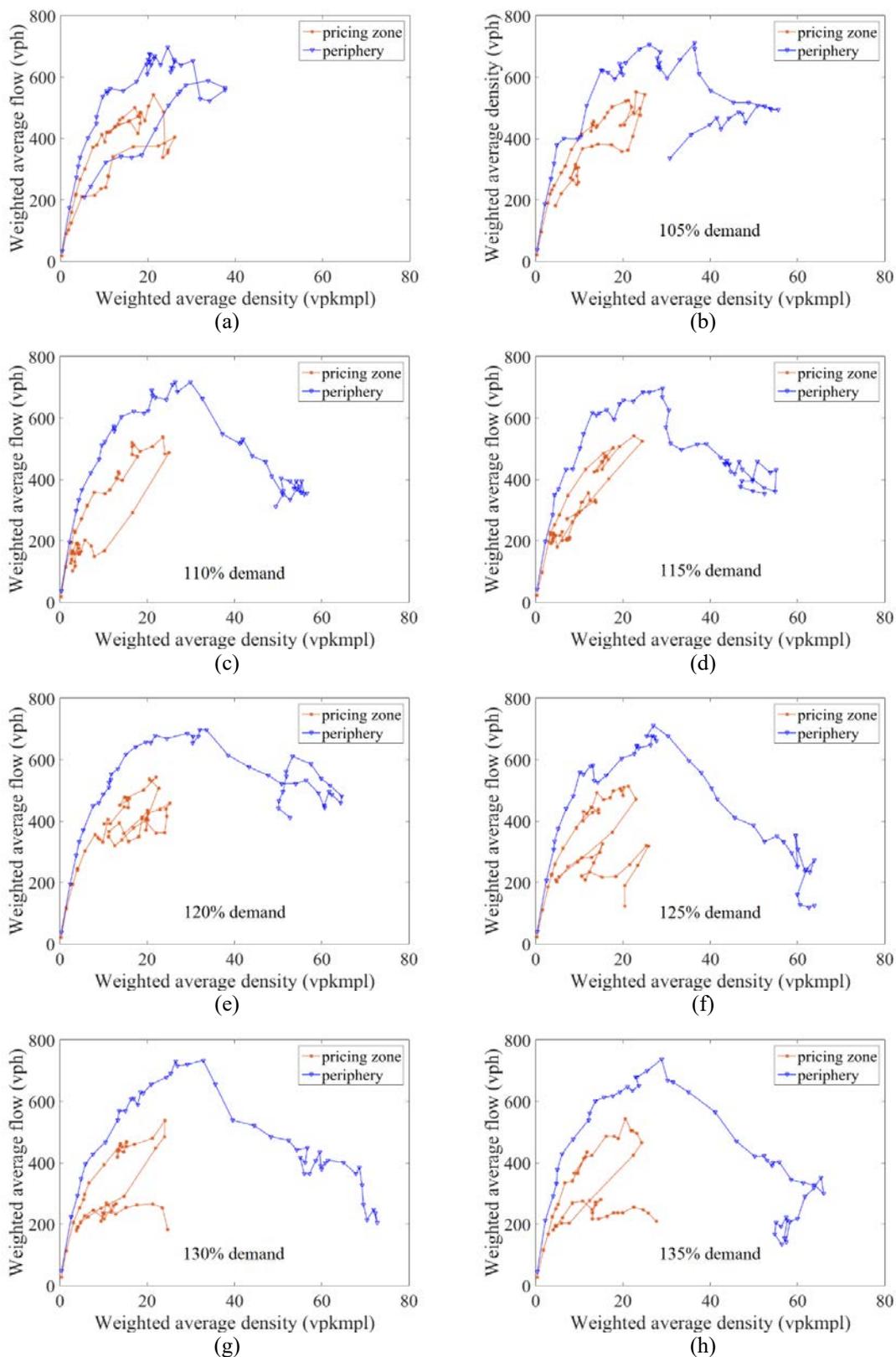

**Fig. 13.** Simulated NFDs of the cordon area and the periphery with incrementally increasing demand.

Depending on the network topology and OD demand, the applicability of the feedback control does vary for different networks. In general, the prerequisite can be satisfied in a real traffic network for two reasons: (i) city ring roads are often available around the urban center for detour traffic which provide enough capacity (e.g. the M1, M2, and M3 highways surrounding the city center of Melbourne); and (ii) the OD demand needs to climb to a level that is seldom realistic for normal daily traffic (e.g. a 15% increase at least for Melbourne). From a methodological perspective, the applicability of the feedback control does motivate further investigation into a coordinated two-region pricing scheme[7].

## 5. Conclusion and future work

In this paper, we develop an SBO framework to solve the optimal toll level problem. A few alternative pricing models are proposed and compared. The first model is a distance toll that is linearly related to the distance traveled within the cordon. While keeping the network from entering the congested regime of the NFD, the distance toll naturally drives users into the shortest paths within the cordon resulting in an uneven distribution of congestion. This is reflected by the large clockwise hysteresis loop in the NFD and the large anticlockwise hysteresis loop in the spread-accumulation relationship. Since the hysteresis loop is an effect of network instability that causes network unproductivity, we aim to reduce this negative effect by proposing a JDTT and JDDT. The JDTT/JDDT is linearly related to the distance traveled and time spent/delay experienced within the cordon. Results show that (i) all the studied pricing models are effective in reducing congestion within the cordon against simulation stochasticity; and (ii) unlike the distance toll, the JDTT and JDDT reduce the uneven distribution of congestion within the cordon, thereby increasing the network stability and productivity. While the global convergence of the pricing control is shown throughout the analysis, further investigation reveals the existence of a prerequisite for applying the feedback control.

The hysteresis loop in the NFD represents a heterogeneous distribution of congestion in the network. The majority of the perimeter control studies assumed a well-defined NFD without hysteresis modeled by a polynomial function (Geroliminis et al., 2013). This assumption does not necessarily hold in a real-world network. While network partitioning can alleviate the issue, there is a need for methodologies that can inherently take into account the natural existence of hysteresis in the NFD (Ramezani et al., 2015). The developed SBO framework for pricing does not rely on any assumptions about the shape of the NFD.

While the proposed pricing schemes have not been implemented anywhere in the world, they certainly point to a promising direction for designing a more efficient and equitable pricing system. Technologies required are no longer an obstacle. Back in 1993, a successful congestion metering trial was undertaken in Cambridge where an on-board metering device was used to monitor vehicle speed and distance (Ison and Rye, 2005). Such a device is also currently used in Oregon, USA to calculate the distance charge. While the privacy concern may be raised, it can be addressed by configuring the device not to transmit location data at all, or when a vehicle is only a few kilometers away from its origins and destinations. This type of practice was previously adopted in Melbourne's road user pricing trial (Transurban, 2016).

A few directions for future research can be considered. With respect to the simulation model, further investigation is needed to relax the inelastic demand and to extend the car traffic to a multi-modal network where public transport is explicitly modeled. A few recent studies (Zheng et al., 2016; Zockaie et al., 2015) have integrated an activity-based model (ABM) with a DTA model to analyze the effect of pricing. Also, it would be more realistic and practical to include user heterogeneity in the model through introduction of multi-class users or a continuously distributed VOT (Jiang and Mahmassani, 2013). With respect to the pricing models, further investigation is needed to propose a coordinated pricing scheme for a two- or multi-region network. Whether there is any way by which we can explicitly optimize the balance between $\alpha$ and $\beta_1$ remains an open question. The answer is perhaps, yes. A possible way is through bi-objective optimization where we introduce a secondary objective for which the balance is to be optimized. Another possible way is to apply surrogate modeling as previously mentioned instead of using the PI controller. We are currently working towards this direction. An interesting question also arises about the effect of users' adaptivity on pricing (Saberi et al., 2014b). With the development of information and communication technologies and potential impacts of connected and automated vehicles, a high level of users' adaptivity can be realized.

---

[7] See Geroliminis et al. (2013) for a coordinated two-region perimeter control scheme.

**Appendix A. Derivation of the discrete PI controller**

The parallel form of the continuous-time PI controller is expressed as follows:

$$u(t) = P_\text{p} \cdot e(t) + P_\text{i} \cdot \int_0^t e(\tau)\, d\tau \tag{A.1}$$

where $u(t)$ is the control input as a function of time $t$; $e(t)$ is the error between the set point and the measured output as a function of time $t$; $P_\text{p}$ and $P_\text{i}$ are positive proportional and integral gain parameters, respectively. The Laplace transform of Eq. (A.1) is provided below:

$$\frac{U(s)}{E(s)} = P_\text{p} + \frac{P_\text{i}}{s} \tag{A.2}$$

where $s$ is a complex number. Since the pricing controller is not real-time, we perform the z-transform on Eq. (A.2) using the backward Euler method and obtain the following discrete-time equivalent:

$$\frac{U(z)}{E(z)} = P_\text{p} + \frac{P_\text{i} \cdot T_\text{s}}{1 - z^{-1}} \tag{A.3}$$

where $T_\text{s}$ is the sampling period and $z$ is a complex number. Without loss of generality, we consider $P_\text{i} \cdot T_\text{s}$ as the new $P_\text{i}$ and perform the inverse z-transform on Eq. (A.3):

$$(1 - z^{-1}) \cdot U(z) = P_\text{p} \cdot (1 - z^{-1}) \cdot E(z) + P_\text{i} \cdot E(z) \tag{A.4}$$

$$u(i) = u(i-1) + P_\text{p} \cdot \big(e(i) - e(i-1)\big) + P_\text{i} \cdot e(i) \tag{A.5}$$

Given $\alpha$ and $e(i) = K_\text{max}(i) - K_\text{cr}$, Eq. (A.5) is rewritten as follows:

$$\alpha(i) = \alpha(i-1) + P_\text{p} \cdot \big(K_\text{max}(i) - K_\text{max}(i-1)\big) + P_\text{i} \cdot (K_\text{max}(i) - K_\text{cr}) \tag{A.6}$$

Interested readers may refer to Aboudolas and Geroliminis (2013) for an alternative way to derive the discrete PI controller.

**Appendix B. Proof of Proposition 1**

We expand Eq. (24) recursively as follows:

$$\boldsymbol{u}(i^*) = \boldsymbol{u}(i^*-1) + \boldsymbol{\mu P E}(i^*) \tag{B.1}$$

$$= \boldsymbol{u}(i^* - 2) + \sum_{j=i^*-1}^{i^*} \boldsymbol{\mu PE}(j)$$

$$= \cdots$$

$$= \boldsymbol{u}(1) + \sum_{j=2}^{i^*} \boldsymbol{\mu PE}(j)$$

We calculate the summation of matrix multiplications:

$$\sum_{j=2}^{i^*} \boldsymbol{\mu PE}(j) = \sum_{j=2}^{i^*} \begin{bmatrix} \mu_\alpha & 0 \\ 0 & \mu_{\beta_1} \end{bmatrix} \begin{bmatrix} P_p & P_i \\ P_p & P_i \end{bmatrix} \begin{bmatrix} E_p(j) \\ E_i(j) \end{bmatrix}$$

$$= \sum_{j=2}^{i^*} \begin{bmatrix} \mu_\alpha \cdot P_p \cdot E_p(j) + \mu_\alpha \cdot P_i \cdot E_i(j) \\ \mu_{\beta_1} \cdot P_p \cdot E_p(j) + \mu_{\beta_1} \cdot P_i \cdot E_i(j) \end{bmatrix} \quad \text{(B.2)}$$

$$= \begin{bmatrix} \mu_\alpha \cdot \sum_{j=2}^{i^*} [P_p \cdot E_p(j) + P_i \cdot E_i(j)] \\ \mu_{\beta_1} \cdot \sum_{j=2}^{i^*} [P_p \cdot E_p(j) + P_i \cdot E_i(j)] \end{bmatrix}$$

Based on Eqs. (25) and (B.2), Eq. (B.1) is rewritten as follows which completes the proof:

$$\boldsymbol{u}(i^*) = \begin{bmatrix} \alpha(i^*) \\ \beta_1(i^*) \end{bmatrix} = \begin{bmatrix} \mu_\alpha \cdot \left\{ P_i \cdot E_i(1) + \sum_{j=2}^{i^*} [P_p \cdot E_p(j) + P_i \cdot E_i(j)] \right\} \\ \mu_{\beta_1} \cdot \left\{ P_i \cdot E_i(1) + \sum_{j=2}^{i^*} [P_p \cdot E_p(j) + P_i \cdot E_i(j)] \right\} \end{bmatrix} \quad \text{(B.3)}$$

**Appendix C. Simulation model of Melbourne, Australia**

The calibrated DTA model of Melbourne, Australia is deployed in AIMSUN as a discrete-event lane-based simulation (Shafiei et al., 2017). Each link has information about its geometry including the number of lanes, capacity, speed limit and parameters of the traffic flow fundamental diagram. Each node is modeled as a queue server. While being capable of replicating traffic dynamics including queue spillbacks, mesoscopic simulation largely eases the computational complexity of simulating large-scale dynamic traffic networks compared with microscopic simulation (see TSS (2014) for further details).

The network configuration of the Melbourne metropolitan area is obtained from the Victorian Integrated Transport Model (VITM). Figure C. 1 shows the extracted sub-network from the greater Melbourne area model and the time-dependent OD demand. Table C. 1 summarizes the network topology. The boundary of the sub-network is marked by the dash lines and the inner rectangle represents the pricing zone. The selected cordon does not necessarily represent the optimal one for pricing. Further investigation is underway to formulate an optimal toll area problem. In the simulation model, the DTA function is active across the entire network. Signal controls at intersections are set as actuated signals using the Sydney Coordinated Adaptive Traffic System (SCATS) data including the maximum cycle time, minimum green time, and turning movements for each phase. While traffic flow fundamental diagrams are calibrated against freeway loop detector data from multiple months (Gu et al., 2016, 2017), the time-dependent OD demand is calibrated and validated using multi-source traffic data (Shafiei et al., 2018).

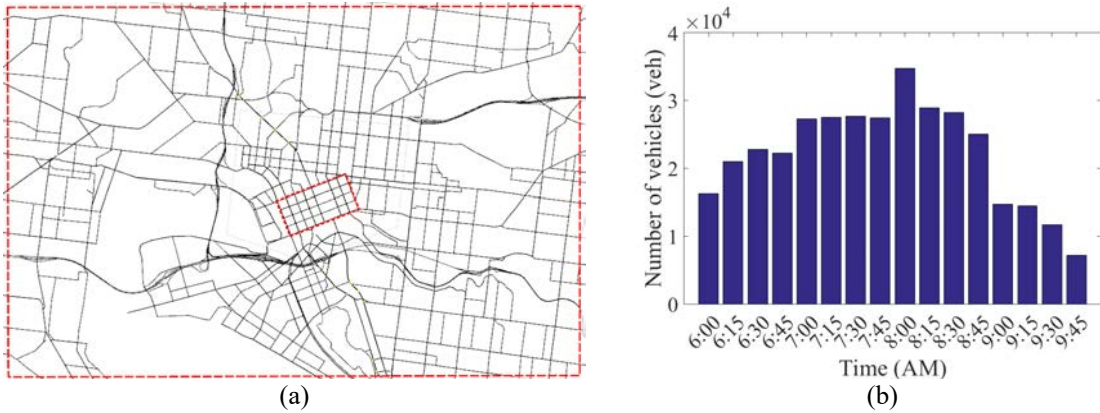

(a) (b)

**Figure C. 1.** (a) The extracted sub-network from the greater Melbourne area model; (b) time-dependent OD demand.

**Table C. 1** Network topology of the extracted sub-network.

| | |
|---|---|
| Total number of links | 4,375 |
| Total number of nodes | 1,977 |
| Total number of centroids | 492 |
| Number of links in the pricing zone | 282 |
| Number of nodes in the pricing zone | 91 |
| Number of centroids in the pricing zone | 30 |

**Appendix D. Results of the non-tolling and optimal cordon toll scenarios**

The maximum network flow without pricing occurs when the network density is between 20 and 30 vpkmpl. We therefore set $K_{cr} = 25$ and determine the tolling period to be 40 minutes from 8:35 to 9:15 AM. We apply the developed SBO framework and obtain an optimal cordon toll of around $1.9. The simulated NFDs and time series of the average network density are shown in Figure. D. 1(a) and (b), respectively. The cordon toll successfully keeps the network from entering the congested regime of the NFD and reduces the size of the hysteresis loop. This is because a proportion of users are priced off the cordon area resulting in a lower level of congestion and hence a more homogenous distribution of congestion.

Analytical derivation of the gain parameters in the PI controller is intractable due to complex system dynamics. A sensitivity analysis is therefore presented to (i) verify the global convergence of the feedback control; and (ii) provide general guidance on using the trial-and-error method. Three pairs of gain parameters are tested and the results are shown in Figure. D. 1(c). Although different values of $P_p$ and $P_i$ are used, the optimal cordon tolls have a unique convergent solution suggesting that the global convergence law holds. Nevertheless, when $P_p = 0.4$ and $P_i = 0.05$, the oscillatory behavior of the PI controller becomes significant making it difficult to pinpoint the optimal solution. Once the gain parameters are lowered, the oscillation becomes less severe although at the cost of an increased number of iterations until convergence. Given this trade-off, a general principle of applying the trial-and-error method is to start with a slightly larger pair of $P_p$ and $P_i$ and gradually decrease their values until a relatively smooth convergence pattern is achieved.

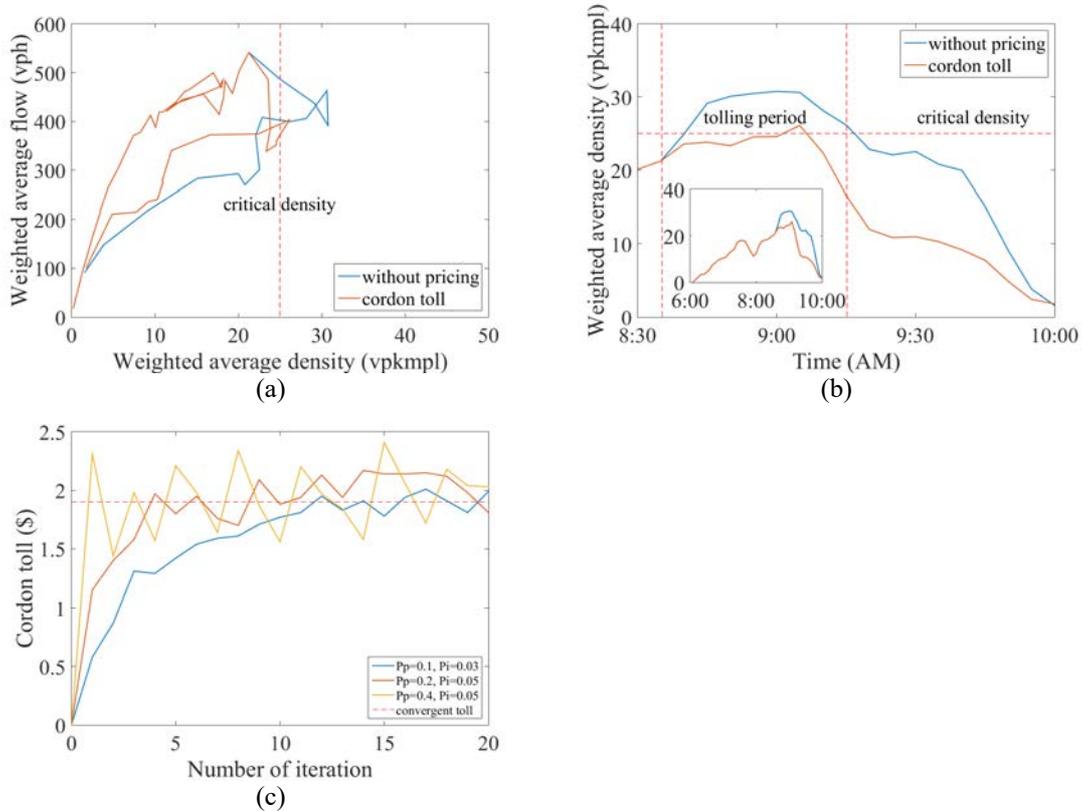

**Figure. D. 1.** Simulation results of the cordon area under the non-tolling and optimal cordon toll scenarios: (a) simulated NFDs; (b) time series of the average network density; (c) a sensitivity analysis on the controller gain parameters suggesting a unique convergent solution.

## Appendix E. Obtaining the lower envelop in the spread-accumulation relationship

We assume a third-order polynomial function and perform ten simulation runs without pricing to estimate the coefficients. Since the fitted function only serves as a mathematical approximation to better represent the lower envelope, it may not necessarily be the best functional form (Simoni et al., 2015). The fitted lower envelope is shown in Figure. E. 1 and mathematically expressed as $\gamma(k) = -0.0003154 \cdot k^3 + 0.01499 \cdot k^2 + 1.127 \cdot k$.

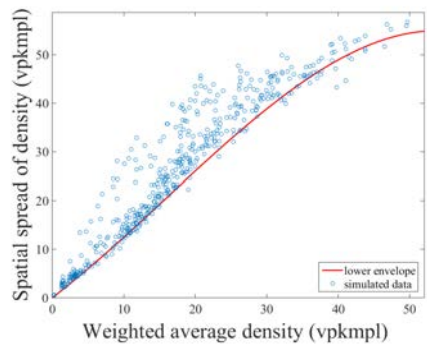

**Figure. E. 1.** The fitted polynomial function for the lower envelop in the spread-accumulation relationship.

**Appendix F. Simulated NFDs of the cordon area and periphery under different pricing scenarios**

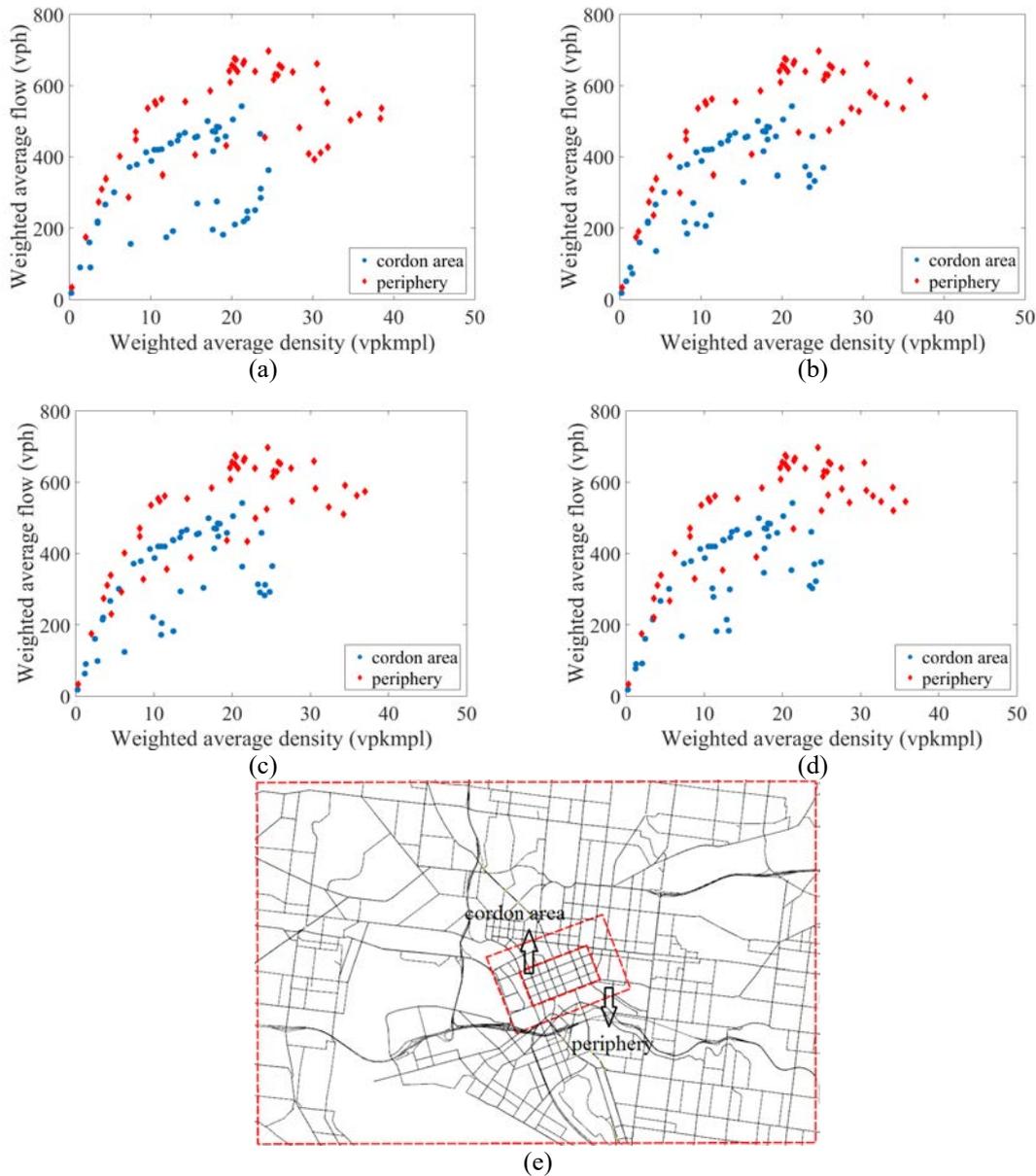

**Figure. F. 1.** Simulated NFDs of the cordon area and periphery under different pricing scenarios: (a) distance toll; (b) JDTT simultaneous; (c) JDTT sequential; (d) JDDT sequential; (e) an illustration of the cordon area and its periphery.

**References**

Aboudolas, K., Geroliminis, N., 2013. Perimeter and boundary flow control in multi-reservoir heterogeneous networks. *Transp. Res. Part B* 55, 265-281.

Arnott, R., De Palma, A., Lindsey, R., 1990. Economics of a bottleneck. *J. Urban Econ* 27(1), 111-130.

Arnott, R., De Palma, A., Lindsey, R., 1993. A structural model of peak-period congestion: A traffic bottleneck with elastic demand. *Am. Econ. Rev.*, 161-179.

Buisson, C., Ladier, C., 2009. Exploring the Impact of Homogeneity of Traffic Measurements on the Existence of Macroscopic Fundamental Diagrams. *Transp. Res. Rec.: J. Transp. Res. Board*(2124), 127-136.

Cascetta, E., 2001. *Transportation Systems Engineering: Theory and Methods*. Kluwer Academic Publishers, Dordrecht, the Netherlands.

Cascetta, E., Nuzzolo, A., Russo, F., Vitetta, A., 1996. A modified logit route choice model overcoming path overlapping problems. Specification and some calibration results for interurban networks, *Proceedings of the 13th Internaional Symposium on Transportation and Traffic Theory*, Lyon, France, pp. 697-711.

Chen, X., Xiong, C., He, X., Zhu, Z., Zhang, L., 2016. Time-of-day vehicle mileage fees for congestion mitigation and revenue generation: A simulation-based optimization method and its real-world application. *Transp. Res. Part C* 63, 71-95.

Chen, X., Zhang, L., He, X., Xiong, C., Li, Z., 2014. Surrogate-Based Optimization of Expensive-to-Evaluate Objective for Optimal Highway Toll Charges in Transportation Network. *Comput.-Aided Civ. Inf. Eng.* 29(5), 359-381.

Daganzo, C.F., 2007. Urban gridlock: Macroscopic modeling and mitigation approaches. *Transp. Res. Part B* 41(1), 49-62.

Daganzo, C.F., Lehe, L.J., 2015. Distance-dependent congestion pricing for downtown zones. *Transp. Res. Part B* 75, 89-99.

De Palma, A., Lindsey, R., 2011. Traffic congestion pricing methodologies and technologies. *Transp. Res. Part C* 19(6), 1377-1399.

Ekström, J., Kristoffersson, I., Quttineh, N.-H., 2016. Surrogate-based optimization of cordon toll levels in congested traffic networks. *J. Adv. Transp.* 50(6), 1008-1033.

Ekström, J., Sumalee, A., Lo, H.K., 2012. Optimizing toll locations and levels using a mixed integer linear approximation approach. *Transp. Res. Part B* 46(7), 834-854.

Gayah, V.V., Daganzo, C.F., 2011. Clockwise hysteresis loops in the Macroscopic Fundamental Diagram: An effect of network instability. *Transp. Res. Part B* 45(4), 643-655.

Geroliminis, N., Daganzo, C.F., 2008. Existence of urban-scale macroscopic fundamental diagrams: Some experimental findings. *Transp. Res. Part B* 42(9), 759-770.

Geroliminis, N., Haddad, J., Ramezani, M., 2013. Optimal perimeter control for two urban regions with macroscopic fundamental diagrams: A model predictive approach. *IEEE Trans. Intell. Transp. Syst.* 14(1), 348-359.

Geroliminis, N., Levinson, D., 2009. Cordon Pricing Consistent with the Physics of Overcrowding, in: Lam, W.H.K., Wong, S.C., Lo, H.K. (Eds.), *Transportation and Traffic Theory 2009: Golden Jubilee*. Springer US, pp. 219-240.

Godfrey, J.W., 1969. The mechanism of a road network. *Traffic Eng. Control* 11(7), 323-327.

Gu, Z., Liu, Z., Cheng, Q., Saberi, M., 2018. Congestion pricing practices and public acceptance: A review of evidence. *Case Stud. Transp. Policy* 6(1), 94-101.

Gu, Z., Saberi, M., Sarvi, M., Liu, Z., 2016. Calibration of Traffic Flow Fundamental Diagrams for Network Simulation Applications: A Two-Stage Clustering Approach, *Proceedings of the 19th International IEEE Conference on Intelligent Transportation Systems (ITSC)*, Rio de Janeiro, Brazil, pp. 1348-1353.

Gu, Z., Saberi, M., Sarvi, M., Liu, Z., 2017. A big data approach for clustering and calibration of link fundamental diagrams for large-scale network simulation applications. *Transp. Res. Part C*, in press.

Haddad, J., Shraiber, A., 2014. Robust perimeter control design for an urban region. *Transp. Res. Part B* 68, 315-332.

He, X., Chen, X., Xiong, C., Zhu, Z., Zhang, L., 2017. Optimal Time-Varying Pricing for Toll Roads Under Multiple Objectives: A Simulation-Based Optimization Approach. *Transp. Sci.* 51(2), 412-426.

Ison, S., Rye, T., 2005. Implementing road user charging: the lessons learnt from Hong Kong, Cambridge and Central London. *Transp. Rev.* 25(4), 451-465.

Ji, Y., Geroliminis, N., 2012. On the spatial partitioning of urban transportation networks. *Transp. Res. Part B* 46(10), 1639-1656.

Jiang, L., Mahmassani, H., 2013. Toll Pricing: Computational Tests for Capturing Heterogeneity of User Preferences. *Transp. Res. Rec.: J. Transp. Res. Board*(2343), 105-115.

Keyvan-Ekbatani, M., Kouvelas, A., Papamichail, I., Papageorgiou, M., 2012. Exploiting the fundamental diagram of urban networks for feedback-based gating. *Transp. Res. Part B* 46(10), 1393-1403.

Keyvan-Ekbatani, M., Papageorgiou, M., Knoop, V.L., 2015. Controller Design for Gating Traffic Control in Presence of Time-delay in Urban Road Networks. *Transp. Res. Procedia* 7, 651-668.

Knight, F.H., 1924. Some fallacies in the interpretation of social cost. *Q. J. Econ.*, 582-606.

Knoop, V., Hoogendoorn, S., 2013. Empirics of a Generalized Macroscopic Fundamental Diagram for Urban Freeways. *Transp. Res. Rec.: J. Transp. Res. Board*(2391), 133-141.

Kosmatopoulos, E.B., Papageorgiou, M., 2003. Stability analysis of the freeway ramp metering control strategy ALINEA, *Proceedings of the 11th IEEE Mediterranean Conference on Control and Automation*, Rhodes, Greece.

Leclercq, L., Chiabaut, N., Trinquier, B., 2014. Macroscopic Fundamental Diagrams: A cross-comparison of estimation methods. *Transp. Res. Part B* 62, 1-12.

Lee, M., 2018. Getting Around Metro Vancouver: A Closer Look at Mobility Pricing and Fairness, Vancouver, Canada.

Legaspi, J., Douglas, N., 2015. Value of Travel Time Revisited–NSW Experiment, *Proceedings of the 37th Australasian Transport Research Forum (ATRF)*, Sydney, NSW, Australia.

Li, M.Z., 2002. The role of speed–flow relationship in congestion pricing implementation with an application to Singapore. *Transp. Res. Part B* 36(8), 731-754.

Liu, L.N., McDonald, J.F., 1999. Economic efficiency of second-best congestion pricing schemes in urban highway systems. *Transp. Res. Part B* 33(3), 157-188.

Liu, Z., Meng, Q., Wang, S., 2013. Speed-based toll design for cordon-based congestion pricing scheme. *Transp. Res. Part C* 31, 83-98.

Liu, Z., Wang, S., Meng, Q., 2014. Optimal joint distance and time toll for cordon-based congestion pricing. *Transp. Res. Part B* 69, 81-97.

Mahmassani, H., Williams, J.C., Herman, R., 1987. Performance of urban traffic networks, *Proceedings of the 10th International Symposium on Transportation and Traffic Theory*. Elsevier, pp. 1-20.

Mahmassani, H.S., Saberi, M., Zockaie, A., 2013. Urban network gridlock: Theory, characteristics, and dynamics. *Transp. Res. Part C* 36, 480-497.

May, A.D., Milne, D.S., 2000. Effects of alternative road pricing systems on network performance. *Transp. Res. Part A* 34(6), 407-436.

Mazloumian, A., Geroliminis, N., Helbing, D., 2010. The spatial variability of vehicle densities as determinant of urban network capacity. *Philos. Trans. Roy. Soc. A* 368(1928), 4627-4647.

Meng, Q., Liu, Z., Wang, S., 2012. Optimal distance tolls under congestion pricing and continuously distributed value of time. *Transp. Res. Part E* 48(5), 937-957.

Olszewski, P., Fan, H.S., Tan, Y.-W., 1995. Area-wide traffic speed-flow model for the Singapore CBD. *Transp. Res. Part A* 29(4), 273-281.
Osorio, C., Bierlaire, M., 2013. A simulation-based optimization framework for urban transportation problems. *Oper. Res.* 61(6), 1333-1345.
Osorio, C., Chong, L., 2015. A computationally efficient simulation-based optimization algorithm for large-scale urban transportation problems. *Transp. Sci.* 49(3), 623-636.
Papageorgiou, M., Hadj-Salem, H., Blosseville, J.-M., 1991. ALINEA: A local feedback control law for on-ramp metering. *Transp. Res. Rec.: J. Transp. Res. Board* 1320(1), 58-67.
Papageorgiou, M., Kotsialos, A., 2002. Freeway ramp metering: an overview. *IEEE Trans. Intell. Transp. Syst.* 3(4), 271-281.
Pierce, G., Shoup, D., 2013. Getting the prices right: an evaluation of pricing parking by demand in San Francisco. *J. Am. Plan. Assoc.* 79(1), 67-81.
Pigou, A.C., 1920. *The Economics of Welfare*. MacMillan, London.
Ramezani, M., Haddad, J., Geroliminis, N., 2015. Dynamics of heterogeneity in urban networks: aggregated traffic modeling and hierarchical control. *Transp. Res. Part B* 74, 1-19.
Saberi, M., Mahmassani, H., 2012. Exploring Properties of Network-wide Flow-Density Relations in A Freeway Network. *Transp. Res. Rec.: J. Transp. Res. Board*(2315), 153-163.
Saberi, M., Mahmassani, H., 2013. Hysteresis and capacity drop phenomena in freeway networks: Empirical characterization and interpretation. *Transp. Res. Rec.: J. Transp. Res. Board*(2391), 44-55.
Saberi, M., Mahmassani, H.S., Hou, T., Zockaie, A., 2014a. Estimating Network Fundamental Diagram Using Three-Dimensional Vehicle Trajectories: Extending Edie's Definitions of Traffic Flow Variables to Networks. *Transp. Res. Rec.: J. Transp. Res. Board*(2422), 12-20.
Saberi, M., Mahmassani, H.S., Zockaie, A., 2014b. Network capacity, traffic instability, and adaptive driving: findings from simulated urban network experiments. *EURO J. Transp. Logist.* 3(3-4), 289-308.
Saeedmanesh, M., Geroliminis, N., 2016. Clustering of heterogeneous networks with directional flows based on "Snake" similarities. *Transp. Res. Part B* 91, 250-269.
Saeedmanesh, M., Geroliminis, N., 2017. Dynamic clustering and propagation of congestion in heterogeneously congested urban traffic networks. *Transp. Res. Part B* 105, 193-211.
Shafiei, S., Gu, Z., Saberi, M., 2018. Calibration and Validation of a Simulation-based Dynamic Traffic Assignment Model for a Large-Scale Congested Network. *Simul. Modell. Pract. Theory*, accepted.
Shafiei, S., Gu, Z., Sarvi, M., Saberi, M., 2017. Deployment and Calibration of a Large-Scale Mesoscopic Dynamic Traffic Assignment Model of Melbourne, Australia, *Transportation Research Board 96th Annual Meeting*, Washington, DC.
Simoni, M., 2013. Congestion pricing schemes controlled by the gMFD: a comprehensive design and appraisal to bridge the engineering and economic perspective (master's thesis). Delft University of Technology, Delft, the Netherlands.
Simoni, M.D., Pel, A.J., Waraich, R.A., Hoogendoorn, S.P., 2015. Marginal cost congestion pricing based on the network fundamental diagram. *Transp. Res. Part C* 56, 221-238.
Transurban, 2016. Changed Conditions Ahead: The Transport Revolution and What it Means for Australians, Melbourne, Australia.
TSS, 2014. Aimsun 8 Dynamic Simulators Users' Manual, Barcelona, Spain.

Verhoef, E.T., 2002. Second-best congestion pricing in general networks. Heuristic algorithms for finding second-best optimal toll levels and toll points. *Transp. Res. Part B* 36(8), 707-729.
Vickrey, W., 1963. Pricing and resource allocation in transportation and public utilities. *Am. Econ. Rev.* 53(2), 452-465.
Yang, H., Huang, H., 2005. *Mathematical and economic theory of road pricing*. Elsevier, Oxford.
Yang, H., Meng, Q., Lee, D.-H., 2004. Trial-and-error implementation of marginal-cost pricing on networks in the absence of demand functions. *Transp. Res. Part B* 38(6), 477-493.
Yang, H., Zhang, X., 2003. Optimal toll design in second-best link-based congestion pricing. *Transp. Res. Rec.: J. Transp. Res. Board*(1857), 85-92.
Yin, Y., Lou, Y., 2009. Dynamic tolling strategies for managed lanes. *J. Transp. Eng.* 135(2), 45-52.
Zhang, X., Yang, H., 2004. The optimal cordon-based network congestion pricing problem. *Transp. Res. Part B* 38(6), 517-537.
Zheng, N., Geroliminis, N., 2016. Modeling and optimization of multimodal urban networks with limited parking and dynamic pricing. *Transp. Res. Part B* 83, 36-58.
Zheng, N., Rérat, G., Geroliminis, N., 2016. Time-dependent area-based pricing for multimodal systems with heterogeneous users in an agent-based environment. *Transp. Res. Part C* 62, 133-148.
Zheng, N., Waraich, R.A., Axhausen, K.W., Geroliminis, N., 2012. A dynamic cordon pricing scheme combining the Macroscopic Fundamental Diagram and an agent-based traffic model. *Transp. Res. Part A* 46(8), 1291-1303.
Zockaie, A., Mahmassani, H., Saberi, M., Verbas, Ö., 2014. Dynamics of urban network traffic flow during a large-scale evacuation. *Transp. Res. Rec.: J. Transp. Res. Board*(2422), 21-33.
Zockaie, A., Saberi, M., Mahmassani, H.S., Jiang, L., Frei, A., Hou, T., 2015. Towards Integrating an Activity-Based Model with Dynamic Traffic Assignment Considering Heterogeneous User Preferences and Reliability Valuation: Application to Toll Revenue Forecasting in Chicago. *Transp. Res. Rec.: J. Transp. Res. Board*(2493), 78-87.